\pdfoutput=1

\documentclass[11pt]{article}

\usepackage[preprint]{acl}

\usepackage{times}
\usepackage{latexsym}

\usepackage[T1]{fontenc}

\usepackage[utf8]{inputenc}

\usepackage{microtype}

\usepackage{inconsolata}
\usepackage{xcolor}
\usepackage{graphicx}
\usepackage{amsfonts}
\usepackage{amsmath}
\usepackage{amssymb}
\usepackage{amsthm}
\usepackage{multirow}
\usepackage{hhline}
\usepackage{array}
\usepackage{caption} 
\usepackage{subcaption} 
\usepackage{algorithm}
\usepackage{algpseudocode}
\usepackage{listings}
\lstdefinestyle{mystyle}{
    basicstyle=\ttfamily\small,
    captionpos=b,          
    breaklines=true,          
}
\usepackage{enumitem}

\usepackage{multirow}
\usepackage{wrapfig,lipsum,booktabs}
\usepackage{tabularx}
\usepackage{color,colortbl}
\usepackage{longtable}
\title{HyQE: Ranking Contexts with Hypothetical Query Embeddings}

\author{
 \textbf{Weichao Zhou\textsuperscript{1}},
 \textbf{Jiaxin Zhang\textsuperscript{2}},
 \textbf{Hilaf Hasson\textsuperscript{2}},
 \textbf{Anu Singh\textsuperscript{2}},
 \textbf{Wenchao Li\textsuperscript{1}}
\\
 \textsuperscript{1}Boston University
 \textsuperscript{2}Intuit AI Research 
\\
\small{
\textsuperscript{1}\tt{\{zwc662,wenchao\}@bu.edu}
\textsuperscript{2}\tt{\{jiaxin\_zhang,hilaf\_hasson,anu\_singh\}@intuit.com}
}
}

\begin{document}

\include{src}
\maketitle

\begin{abstract}
    In retrieval-augmented systems, context ranking techniques are commonly employed to reorder the retrieved contexts based on their relevance to a user query.
    A standard approach is to measure this relevance through the similarity between contexts and queries in the embedding space.
    However, such similarity often fails to capture the relevance. 
    Alternatively, large language models (LLMs) have been used for ranking contexts.
    However, they can  encounter scalability issues when the number of candidate contexts grows and the context window sizes of the LLMs remain constrained.
    Additionally, these approaches require fine-tuning LLMs with domain-specific data. 
    In this work, we introduce a scalable ranking framework that combines embedding similarity and LLM capabilities without requiring LLM fine-tuning. 
    Our framework uses a pre-trained LLM to hypothesize the user query based on the retrieved contexts and ranks the context based on the similarity between the hypothesized queries and the user query. 
    Our framework is efficient at inference time and is compatible with many other retrieval and ranking techniques. 
    Experimental results show that our method improves the ranking performance across multiple benchmarks. 
    The complete code and data are available at \href{https://github.com/zwc662/hyqe}{https://github.com/zwc662/hyqe}
\end{abstract}
\section{Introduction}
Context retrieval plays a crucial role in natural language processing (NLP). 
Standard techniques can efficiently extract relevant information from dedicated databases to address user queries.
These techniques have been driving advancements in search engines, virtual assistants, and other retrieval-augmented systems by enabling precise, real-time responses in real time, and reducing the risk of hallucination \cite{ram2023context,asai2023retrieval}.

Accurately ranking the relevance of the contexts to the user query is a crucial factor in the performance of retrieval-augmented systems \cite{shi2023large}.
Classical retrieval methods such as TF-IDF and BM25  \cite{bm25} rely on lexical similarities to rank contexts.
Recent advancements in embedding models such as BERT \cite{bert,sentence_bert} have enabled the capture of the semantic similarity between texts through dense vector representations.
To improve the zero-shot performance in unseen contexts, Contriever \cite{contriever} and other successive embedding models are trained via contrastive learning techniques.
However, retrieval with these embedding models focuses on similarity, but similarity alone does not always ensure that the context effectively addresses the query.

LLMs have been incorporated to address this issue.
For instance, LLM-based re-rankers \cite{rankgpt,rankzephyr} can determine whether a context addresses a query better than others.
However, those re-rankers require fine-tuning, which demands extensive dedicated datasets and significant computational resources.
Other methods include using LLM to expand the query before retrieval.
HyDE~\cite{hyde}, for instance, utilizes an LLM to generate hypothetical contexts based on the query, subsequently retrieving concrete contexts that are close to these hypothetical contexts in the embedding space. 
However, the LLM must have sufficient background knowledge about the context to be retrieved so that it can generate semantically similar contexts.
Otherwise, the hypothesis space of the generated contexts would be indefinitely large, and the LLM can generate outdated, irrelevant, hallucinated, and even counterfactual contexts \cite{fewshot,mallen2022not}.
We provide an example later in Fig.\ref{fig:causal}(b), where GPT-3.5-turbo generates outdated information that fails to reflect recent developments on a specific topic.

In this paper, we propose a novel context ranking framework. 
Our approach uses an LLM to generate hypothetical queries based on the existing contexts.
It then measures the relevance between the context and a user-given query based on the similarity between the hypothetical queries and the user-given query.
While our method does not require the LLM to have prior knowledge about the query or the context, the hallucination of the LLM is restrained since a context has limited information and can only provide answers to a certain range of queries.
Furthermore, while HyDE has to use an LLM to generate hypothetical contexts online for every input query, our approach allows retrieving previously generated hypothetical queries for future input queries.
Our method also differs from the LLM-based re-ranker in two-fold.
First, our method does not require fine-tuning an LLM. 
Second, our approach uses text embedding for ranking, while an LLM-based re-ranker has to call an LLM to answer the relevance between every input query and context.
However, our method can be used in conjunction with other ranking methods to iteratively refine the ranking of the retrieved contexts.

In addition to introducing our approach, we compare our approach with existing approaches from the theoretical lens. 
We analyze the causality relationship between the queries and contexts within a class of context ranking approaches, identifying their potential issues, such as their susceptibility to spurious causality relationships.
We then show that our approach mitigates these issues by following a variational inference approach.
Our experimental results demonstrate improvements in ranking the retrieved contexts across multiple information retrieval benchmarks while maintaining efficiency and scalability. Our major contribution is listed as follows. 
\begin{itemize}
    \item We propose to use LLMs to generate hypothetical queries and rank contexts by comparing the similarity between input queries and hypothetical queries.
    \item We examine the causal relationships between queries and contexts in existing context ranking methods and develop a variational inference framework for context ranking. 
    \item We evaluate our method in multiple information retrieval benchmarks by combining different embedding models with different LLMs. 
    The results show that our method can improve the ranking accuracy in most of the benchmarks.
\end{itemize}
\section{Related Work}

\noindent{\bf Retrieval-Augmented Systems} have become a focal point in NLP research, enhancing LLMs by accessing broader knowledge bases beyond LLM context windows \cite{lewis2020rag,rag_survey}. 
These systems use information retrieval techniques to fetch relevant contexts from dedicated databases based on user queries, improving performance in tasks requiring extensive context \cite{mialon2023rag}.

\noindent{\bf Information Retrieval Methods}, such as TF-IDF and BM25, rely on lexical similarities to rank contexts \cite{bm25}. 
Recent advancements in embedding models such as BERT \cite{bert, sentence_bert} allow capturing text semantics through dense vector representations \cite{asai2021one}.
Contrastive learning techniques have further improved the zero-shot performance of embedding models such as Contriever \cite{contriever} in unseen contexts by training the models to differentiate between similar and dissimilar contexts \cite{simcse}.

\noindent{\bf Document Expansion and Query Expansion} are classical techniques to improve retrieval quality and have been widely adopted in RAG systems \cite{query2doc}. 
Query expansion, which dates back to \cite{aqe}, typically involves rewriting the query based on labels \cite{queryexpansion}.
When labels are not available, the query can be expanded with generated contexts \cite{liu2022query}.
For instance, HyDE \cite{hyde} uses LLMs to generate hypothetical contexts based on the input query and uses the embeddings of the query and the hypothetical contexts for retrieval.
However, when the LLM lacks knowledge about the query, query expansion can be susceptible to hallucinated or counterfactual content \cite{fewshot}.

Document expansion \cite{nogueira2019document} involves appending each context with a generated query and creating indexes for the expanded context in the database. 
Our framework also generates queries based on the contexts but does not expand the contexts. 
Studies on generating high-quality queries to build synthetic datasets \cite{almeida-matos-2024-exploring} can be helpful for our framework, but that is not the focus of this paper. 

\noindent{\bf Large Language Models (LLMs)}, from the small-size open source models such as Mistral-7b \cite{mistral7b} to the large-size proprietary models such as GPT-4 \cite{gpt4}, are pre-trained on trillions of tokens, exhibiting unparalleled emergent and generalization abilities across tasks \cite{schaeffer2023are}.
LLMs can be fine-tuned to rank the relevancy between contexts and queries \cite{asai2023self,rankgpt,rankzephyr}. 
Although effective, fine-tuning requires significant computational resources and extensive annotated data \cite{msmarco}.
Furthermore, those methods have to face the challenges related to the context window size \cite{longcontext,kaddour2023challenges,child2019generating,gu2023mamba}, as they combine the query and contexts into a single prompt.
Our method does not use LLMs to evaluate the query-context relevancy.

\noindent{\bf Variational Inference} \cite{variational} sits at the core of our proposed framework.
It has been extensively studied across many fields of machine learning \cite{kingma2013auto,hoffman2013stochastic,zhou2022hierarchical,fellows2019virel}.
In this work, we treat queries and contexts as random variables with causal relationships and reformulate the ranking problem as a probability inference problem.
It is widely known that generative models that respect the causality relationships are more robust to distribution shifts because they can avoid learning spurious relationships between random variables \cite{ahuja2021invariance,scholkopf2021toward,lu2022invariant}. 
In this work, we use an LLM to simulate the query-context relationship while avoiding the intervention of prior knowledge, thereby preserving the causal structure.
\section{Background}

A RAG system retrieves information from a document corpus $C=\{c_1, c_2, \ldots, c_i, \ldots \}$ where each $c_i$ is a context.
Assuming that $Q$ is the whole set of user input queries, given an input query $q\in Q$, a retriever returns a ranked list of relevant contexts from $C$. 
The ranking of those contexts can be evaluated by using Normalized Discounted Cumulative Gain (NDCG) \cite{ndcg} which measures the ranking with graded relevance.
After the retrieval step, one or multiple ranking procedures can be adopted to iteratively refine the quality of the ranking.
We assume that each ranking procedure, including that during retrieval, uses a scoring function $r_q: C\rightarrow \mathbb{R}$ to quantify the relevance between any context $c\in C$ and the query $q$. 
We can rank the contexts with this $r_q$, i.e., $\forall c_1,c_2\in C, c_1\preceq c_2 \Leftrightarrow r(q,c_1) \leq r_q(c_2)$. 

A ranker can just target the first $K$ contexts if the contexts have already been ordered in some previous ranking procedure. 
We denote the set that includes the first $K$ contexts as $C_{q,K}\subseteq C$ such that $|C_{q,K}|=K$.
After ranking those contexts, a new scoring function $r_q$ over $C_{q,K}$ is generated.

When using an embedding model for ranking, we use the cosine similarity between query embedding and context embedding to determine the relevance of the query and context. 
We use $E$ to denote the embedding model.
The cosine similarity between a query $q$ and a context $c$ is 
\begin{eqnarray}
    &sim(q, c)=\frac{\langle E(q), E(c)\rangle }{||E(q)||_2\cdot ||E(c)||_2}&\label{eq:sim}
\end{eqnarray}
As a result, given any query $q$, an embedding model-based ranker's scoring function is defined as  $r_q(c)=sim(q, c)$.
 
\begin{figure*}[htp!]
    \centering
    \begin{subfigure}[b]{\textwidth}
         \centering
         \includegraphics[height=4.6cm, width=16cm]{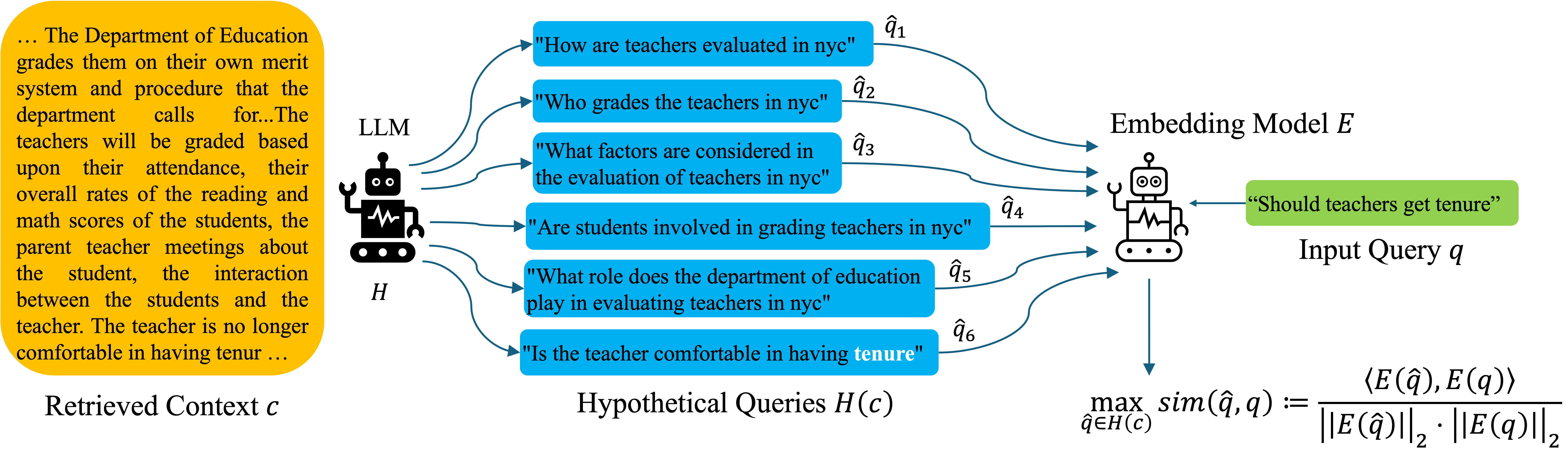}
     \end{subfigure}
 \caption{A flow chart of HyQE ranking framework. Given a query $q$ and a retrieved context $c$, an LLM $H$ is used to generate a set of hypothetical queries $\hat{q}$ from $c$. Then an embedding model $E$ is used to evaluate the semantic similarity between $q$ and ${\hat{q}}$'s. Then cosine similarity is used to determine whether $c$ is relevant to $q$ as in Eq.\ref{eq:hyqe}.
}
\label{fig:diagram}
\end{figure*}

\section{Method}\label{hyqe}
In this section, we introduce our framework for ranking contexts with hypothetical queries.
We first illustrate our context ranking procedure, explain how to obtain those hypothetical queries, and then discuss its complexity.

For each context $c\in C$, we hypothesize the probable queries that the context $c$ can address or the topics it discusses.
We refer to these queries as hypothetical queries, denoted as $\hat{q}$.
For each $c\in C$, we let $H(c)$ denote the set of hypothetical queries associated with $c$.
Our ranking method determines the relevance of a given query $q$ and a context $c$ based on the similarity between the embedding of $q$ and the embedding of $c$, as well as the similarity between those of $q$ and the hypothetical queries $H(c)$ as in Eq.\ref{eq:hyqe} where we introduce a hyperparameter $\lambda$ to balance the two similarities.
\begin{eqnarray}
    &r_q(c) :=   sim(q,c) + \lambda \cdot \underset{\hat{q}\in H(c)}{\max}\ sim(\hat{q},q)\quad\label{eq:hyqe}&
\end{eqnarray}

\noindent{\bf Algorithm \ref{algo:hyqe}} outlines our context ranking procedure.
We start with a set $C_{q,K}$ of $K$ candidate contexts, which are typically the top-$K$ results from a prior ranking step. 
For each context $c\in C_{q,K}$, we generate a set of hypothetical queries ${H}(c)$ by using an LLM, compute the embedding of $c$ and each $\hat{q}\in {H}(c)$ with an embedding model $E$, and then calculate the relevance score $r_q(c)$ using Eq.\ref{eq:sim}.

\begin{algorithm}
\caption{HyQE}\label{algo:hyqe} 
\begin{algorithmic}[1]
\State {\bf Input:} A query $q$; a set $C_{q,K}$ of $K$ candidate contexts; an LLM ${H}$; an embedding model $E$
\State{\bf foreach} context $c$ in $ C_{q,K}$ 
\State \qquad Compute $sim(q, c)$ via Eq.\ref{eq:sim} 
\State \qquad Collect hypothetical queries ${H}(c)$
\State \qquad Compute $r_q(c)$ via Eq.\ref{eq:hyqe}
\State Order $C_{q,K}$ by $r_q(c)$
\State {\bf Output:} The ordered-set $C_{q,K}$
\end{algorithmic}
\end{algorithm}

\noindent{\bf Hypothetical Query Generation.}
Our framework allows utilizing various LLMs ranging from Mistral 7b to GPT-3.5 and GPT-4 to generate hypothetical queries.
Fig.\ref{fig:diagram}(a) shows a flowchart of this query generation process.
For each context $c$, we generate a set of hypothetical queries $H(c)$  by instructing an LLM $H$. 
Specifically, we use a single prompt to generate multiple queries for each context to avoid generating repetitive queries, as shown in Fig.\ref{fig:prompt}.
The prompt is designed to ensure that the generated queries are diverse and relevant to the given context. 
If the length of the context $c$ and the lengths of queries to be generated will exceed the window size of the LLM, we partition $c$, call the LLM to generate queries for one portion at a time, and collect all generated queries in the end.

\begin{figure}
\begin{lstlisting} 
Which kinds of questions can be answered 
based on the following passage

```<passage>
{context}
</passage>'''

Questions must be very short, different, 
and be written on separate lines.
If the passage provides no meaningful 
content, respond with a 'No Content'.
\end{lstlisting}

  \caption{Prompt for hypothetical query generation. `\{context\}' is the placeholder for the context to be filled. }\label{fig:prompt}
\end{figure}

\noindent{\bf Complexity}.
Although generating hypothetical queries for each $c$ with an LLM can be time-consuming, this overhead can be mitigated.
Since the hypothetical queries ${H}(c)$ are independent of the input $q$, once $H(c)$ and the corresponding embeddings are obtained, they can be stored and reused for future queries that involve the same context $c$. 
This eliminates the repetitive query-generation step in line 4 of the algorithm. 
When a previously seen context $c$ is retrieved for some new input query $q'$, we can quickly retrieve the stored $H(c)$  and embeddings of the queries in $H(c)$. 
Then we only need to perform a similarity search to find the hypothetical query $h \in H(c)$ with the closest embedding to the new query $q'$ to compute $r_{q'}(c)$ in line 5. 
By leveraging stored hypothetical queries and their embeddings, our framework ensures efficient and scalable query processing, reducing the computational overhead of real-time LLM calls. 

The complexity of our ranking framework can be broken down as follows.
Generating hypothetical queries $H(c)$ for each context $c\in C$ and computing their embeddings can incur a one-time computational cost.
If each context $c$ can generate $M$ hypothetical queries, the total complexity of this one-time computation is $O(|C|\cdot M)$ where $|C|$ is the total number of contexts and $M$ is limited by the information encompassed in the context.
This complexity is amortized as the number of input queries increases, making our approach more scalable.
If a $c$ is retrieved for a new query $q'$ and its hypothetical query set $H(c)$ has been indexed, the one-time computational complexity of retrieving the closest hypothetical query $\hat{q}\in H(c)$ via similarity search is typically sub-linear in $|H(c)|$. 

In comparison, query expansion methods typically generate contexts for each input query. 
Thus, the total complexity cannot be amortized by the growing number of input queries.
For instance, HyDE \cite{hyde} requires generating a group of hypothetical documents for each query, leading to a total complexity of  $O(|Q| \cdot N)$ where $|Q|$ is the number of queries and $N$ is the number of hypothetical contexts, both of which are independent of the document corpus $C$ and can be infinite. 

Similar comparisons apply to LLM-based re-rankers \cite{rankgpt,rankzephyr}, where the complexity is proportional to the lengths of the input query and the retrieved contexts, as the re-rankers require concatenating the query and contexts in the prompts to generate responses.
This complexity cannot be amortized, making ranking contexts expensive as the number of queries increases, considering that LLM-based re-rankers are often large, proprietary models. 
HyQE allows using small pre-trained LLMs and open-source embedding models, significantly reducing operational costs while maintaining efficiency and effectiveness.
 
\begin{figure*}[ht!]
\centering
    \begin{subfigure}[b]{0.32\textwidth}
         \centering
         \includegraphics[height=4cm, width=4.5cm]{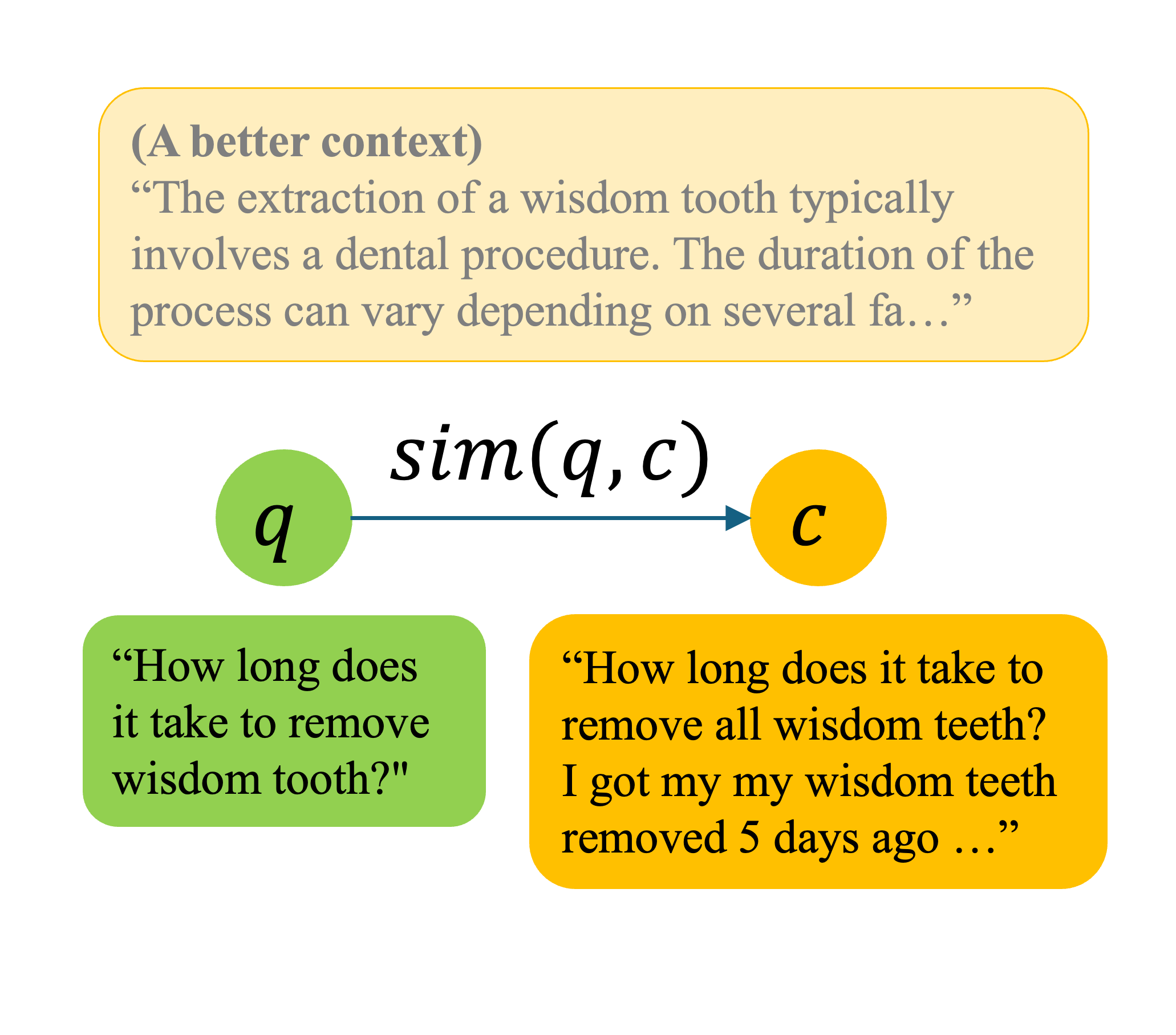}
         \caption{Cosine Similarity} 
     \end{subfigure}
    \begin{subfigure}[b]{0.42\textwidth}
         \centering
         \includegraphics[height=4cm, width=6.8cm]{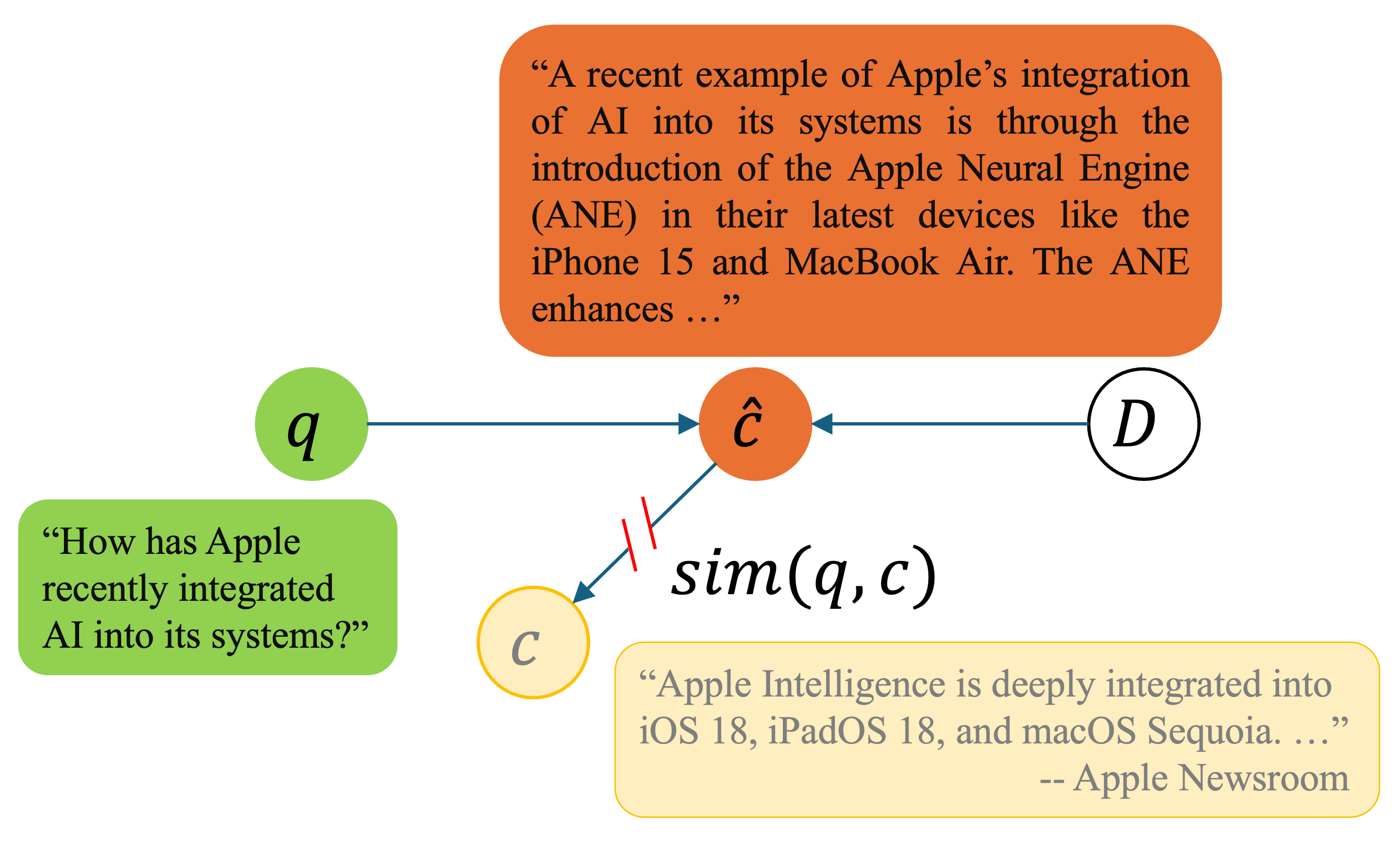}
         \caption{Query Expansion} 
          \end{subfigure}
     \begin{subfigure}[b]{0.24\textwidth}
         \centering
         \includegraphics[height=4cm, width=3.6cm]{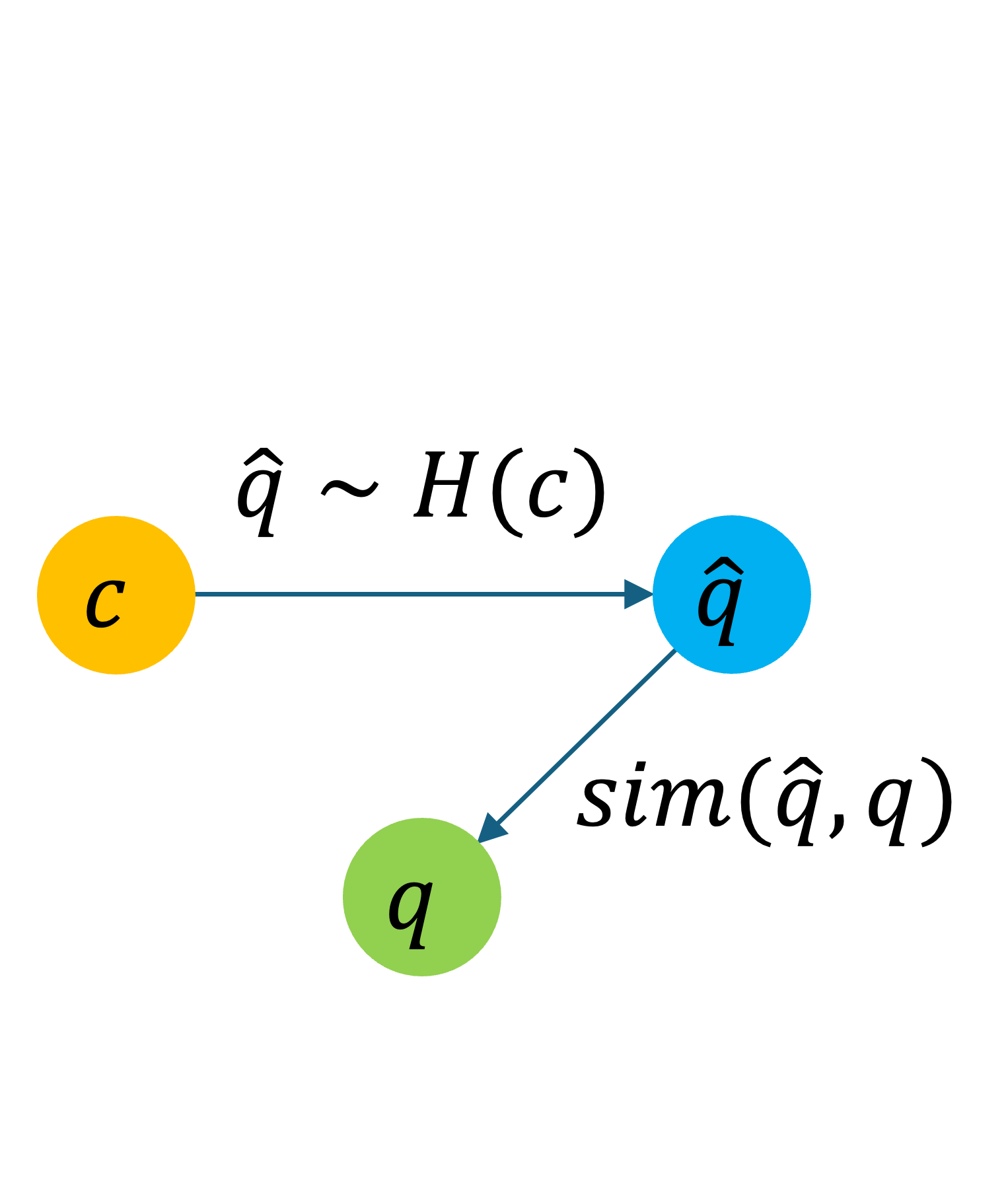}
         \caption{HyQE} 
        \end{subfigure}
        \hfill
        \caption{The random variables $c$ and $q$ respectively indicate context and user input query. (a) Cosine similarity prioritizes semantic similarity rather than retrieving a better context for answering the query. (b) The causality relationship in query expansion methods such as HyDE. The random variable $\hat{c}$ is a hypothetical context, and $D$ indicates the prior knowledge of the LLM used to generate $\hat{c}$. 
        In this example, we use GPT-3.5-turbo to generate a hypothetical context $\hat{c}$ to answer the question in $q$. 
        However, $\hat{c}$ contains outdated information and cannot be used to retrieve the most relevant context $c$ through semantic search.
        (c) The causality relationship in HyQE. An LLM $H$ is used to generate the hypothetical query $\hat{q}$. The causal relationship $q$ and $\hat{q}$ can be simulated with causal similarity.
        } 
    \label{fig:causal}
\end{figure*}

\section{A Variational Inference Perspective}\label{vae}
In this section, we explain how to use variational inference to derive Eq.\ref{eq:hyqe} by establishing the causal relationship between queries and contexts.

\subsection{Causal Relationship between Queries and Contexts}
In Fig.\ref{fig:causal}, we treat context $c$ and query $q$ as two random variables.
We can think of calculating the ranking score $r_q(c)$ as measuring the probability $p(c|q)$ of context $c$ answering question $q$ in the causality model. 
Different ranking methods model the causal relationship between $c$ and $q$ in different ways, resulting in different $p(c|q)$ and  $r_q(c)$.
For instance, the standalone cosine similarity $sim(q,c)$ can produce a spurious $p(c|q)$ since $c$ and $q$ being similar does not necessarily imply that $c$ provides answers to $q$, as shown by the example in Fig.\ref{fig:causal}(a).
Query expansion methods such as HyDE \cite{hyde} introduce a hypothetical context $\hat{c}$ as a latent variable and employ a generative model to simulate $p(\hat{c}|q)$.
However, this causality modeling inevitably involves LLM’s prior knowledge as an intervention \cite{wachter2017counterfactual}, which can lead to spurious causality.
The external knowledge from LLM is represented as an additional variable $D$ from another context space that is indefinitely larger than that of $c$.
As shown in Fig.\ref{fig:causal}(b), it can influence the generation of $\hat{c}$ by introducing outdated, irrelevant, or even counterfactual information \cite{fewshot}.

In contrast, HyQE, as shown in Fig.\ref{fig:causal}(c), introduces a hypothetical query $\hat{q}$ as a latent variable and employs a generative model to simulate $p(\hat{q}|c)$ without involving the prior knowledge of the LLM. 
This confines the generation of hypothetical query $\hat{q}$ strictly within the scope of the context $c$, avoiding the pitfalls of spurious causality and ensuring that the causal relationships remain accurate and relevant. 
This allows us to use cosine similarity to simulate $p(q|\hat{q})$ where $\hat{q}$ and $q$ are both queries.

\subsection{Ranking Contexts from a Variational Inference Perspective}
Now we show how we derive Eq.\ref{eq:hyqe} based on Fig.\ref{fig:causal}(c). 
Given a user query $q$ and a context set $C_{q,K}$, we define $p(c)$ as some prior confidence over the context set $C_{q,K}$ that satisfies $p(c)\propto \exp(sim(q, c))$.  
We let $p(q|c)$ be the probability of context $c$ providing answers to the query $q$, and let $p(q)$ be some prior probability of accepting an input query $q$, which can be seen as a constant when $q$ is already given. 
Based on $p(c)$, $p(q|c)$, and $p(q)$, we aim to learn $p(c|q):= p(c)p(q|c)/ p(q)$, which can be seen as the confidence of the context $c$ addressing the given query $q$. 
Then, we learn $p(c|q)$ by finding a distribution $p_q(c)$ that matches $p(c|q)$ so that we can establish a scoring function based on $p_q(c)$, i.e., $r_q(c)\propto \log p_q(c)$.
This learning objective can be formulated as minimizing the KL-divergence $D_{KL}(p_q(c)|| p(c|q))$ which can be achieved by maximizing the evidence lower-bound (ELBO) of $D_{KL}(p_q(c)|| p(c|q))$ as shown in Eq.\ref{eq:elbo}. 
\begin{eqnarray}
&ELBO(r_q):=\qquad\qquad\qquad\qquad\qquad\qquad & \nonumber\\
&D_{KL}(p_q(c)||p(c)) - \mathbb{E}_{c\sim p_q(c)}[\log p(q|c)]& 
\label{eq:elbo}
\end{eqnarray}
Eq.\ref{eq:elbo} uses a regularization term $D_{KL}(p_q(c)||p(c))$ to penalize $p_q$ if $p_q(c)$ deviates from $p(c)$.
Therefore, we include $sim(q, c)$ as a part of $r_q$ such that the greater $p(c)\propto \exp(sim(q,c))$ is, the greater $p_q(c)\propto\exp (r_q(c))$ becomes.
Meanwhile, the second term in Eq.\ref{eq:elbo} indicates that $p_q$ should also align with $p(q|c)$, the probability of $c$ providing answers to $q$.
To estimate $p(q|c)$, we factorize $\log p(q|c)=\log \mathbb{E}_{\hat{q}\sim p(\hat{q}|c)}[p(q|\hat{q})]$ where $p(\hat{q}|c)$ is the probability of $c$ addressing a hypothetical query $\hat{q}$ and $p(q|\hat{q})$ is the probability of obtaining an input query $q$ given that the semantics of $q$ is equivalent to a given hypothetical query $\hat{q}$.
We can safely use semantic similarity to approximate relevance between queries, i.e., $p(q|\hat{q})\propto \exp(sim(\hat{q}, q))$.
We estimate the expectation w.r.t $p(\hat{q}|c)$ by uniformly sampling from the set $H(c)$ of hypothetical queries such that $\log p(q|c)=\log \mathbb{E}_{h\sim p(\hat{q}|c)}[p(q|\hat{q})]\approx\log \frac{1}{|H(c)|}\sum_{ \hat{q}\in H(c)}p(q| \hat{q})$.
We then have the following two options for further approximation:
 
\noindent{\bf Option 1.} Based on the soft-max approximation, $\log \frac{1}{|H(c)|}\sum_{ \hat{q}\in H(c)} p(q| \hat{q})\approx  \underset{\hat{q}\in H(c)}{\max}\ \log p(q|\hat{q})=\lambda\cdot \underset{ \hat{q}\in H(c)}{\max}\  sim( h, q) + const$ where $\lambda$ is a hyperprameter. 
Then we recover Eq.\ref{eq:hyqe} by ignoring the constant and adding $sim(\hat{q},q)$ mentioned earlier.

\noindent{\bf Option 2.} Based on Jensen's inequality \cite{jensen}, we derive a lower bound of the estimated $\log p(q|c)$ as shown in Eq \ref{eq:jensen},
This allows us to maximize ELBO in Eq.\ref{eq:elbo} by maximizing Eq.\ref{eq:jensen}, resulting in an alternative of Eq.\ref{eq:hyqe} as shown in Eq.\ref{eq:hyqe_alt}.
    \begin{eqnarray}
         &\log \frac{1}{|H(c)|}\sum_{ \hat{q}\in H(c)} p(q| \hat{q})&\nonumber\\
         &\geq\frac{1}{|H(c)|}\sum_{ \hat{q}\in H(c)} \log p(q| \hat{q})&\nonumber\\
        &= \lambda \cdot  \frac{1}{|H(c)|}\sum_{ \hat{q}\in H(c)} sim(q,  \hat{q})  + const\ \label{eq:jensen} &
    \end{eqnarray}
In our HyQE framework, we mainly focus on Option 1. 
We will compare Option 1 with Option 2 in our evaluation.
\begin{eqnarray}
   && r_q(c) :=   sim(q,c) + \nonumber\\
   && \qquad  \lambda \cdot \frac{1}{|H(c)|}\sum_{ \hat{q}\in H(c)} sim(q,  \hat{q})  \label{eq:hyqe_alt} 
\end{eqnarray}

\section{Experiments}\label{sec:exp}
\begin{table*}[!ht]
\centering
\resizebox{1.8\columnwidth}{!}{%
\begin{tabular}{m{6.5em} | m{9.2em} | m{7.7em} | m{2.3em}  m{2.3em} m{2.3em} m{2.3em} m{2.3em}} 
\toprule
Retrieval Model & Embedding Model & HyQE Model  & DL19 & DL20 & COVID & NEWS & Touche \\
\hline
\multirow{4}*{contriever}               & \multirow{4}*{contriever}             &  -   & {\color{red}44.54}         & {\color{red}42.13}        & {\color{red}27.32}        &  {\color{red}34.84}       & {\color{red}16.68} \\ 
                                         \cline{3-8}
                                        &                                       &  GPT-4o & {\color{blue}53.97}   & {\color{blue}51.93}  & {35.03}  &  {41.27} & {17.78} \\  
                                        \cline{3-8}
                                        &                                       &  GPT-3.5-turbo & {53.19}   & {50.04}  & {35.06}  &  {42.33} & {\color{blue}21.02} \\  
                                        \cline{3-8}
                                        &                                       &  Mistral-7b-instruct & {52.28}  & {49.62}  & {\color{blue}35.54}  & {\color{blue}42.56} & {20.78} \\  
\hline
\multirow{4}*{bge-base-en-v1.5}         & \multirow{4}*{bge-base-en-v1.5}  & -   & {\color{red}70.39} & {\color{red}68.30} & {\color{red}69.96}  & {\color{red}40.94} & {\color{red}18.99} \\ 
                                         \cline{3-8}
                                        &                                       &  GPT-4o & {\color{blue}72.04}   & {\color{blue}69.42}  & {\color{blue}80.29}  &  {43.01} & {19.44} \\  
                                        \cline{3-8}
                                        &                                       &  GPT-3.5-turbo &  {71.77} & {68.33} & {80.13}  & {\color{blue}44.03} & {20.14} \\  
                                        \cline{3-8}
                                        &                                       &  Mistral-7b-instruct & {70.72}  & {69.02}  & {78.93}  & {43.34} & {\color{blue}21.36} \\ 
\hhline{=|=|=|= = = = =}
\multirow{16}*{SPLADE++ ED}                
                                        & \multirow{3}*{contriever}             &       -        & {\color{red}53.47} & {\color{red}53.51} & { 67.35} & {\color{red}39.01} & {\color{red}20.45} \\
                                         \cline{3-8}
                                        &                                       &  GPT-4o & {\color{blue}60.68}   & {\color{blue}61.66}  & {64.90}  &  {44.45} & {19.17} \\  
                                        \cline{3-8}
                                        &                                       & GPT-3.5-turbo  & {60.08} & {58.27} & {65.97} & {\color{blue}44.79} & {\color{blue}23.01} \\   
                                        \cline{3-8}
                                        &                                       &  Mistral-7b-instruct    & {57.99}  & {59.59}  & {65.78}  & {44.33} & {22.32} \\ 
                                        \cline{2-8}
                                        & \multirow{3}*{bge-base-en-v1.5}       &       -        & {\color{red}71.25} & {\color{red}68.58} & {\color{red}80.45} & {46.21} & {\color{red}21.53} \\  
                                        \cline{3-8}
                                        &                                       &  GPT-4o & {\color{blue}72.35}   & {68.96}  & {80.82}  &  {\color{blue}46.25} & {22.11} \\  
                                        \cline{3-8}
                                        &                                       &  GPT-3.5-turbo & {71.66} & {68.83} & {\color{blue}81.55} & {46.18} & {\color{blue}23.15} \\ 
                                        \cline{3-8}
                                        &                                       &  Mistral-7b-instruct    & {71.78}  & {\color{blue}69.06}  & {80.82}  & {45.97} & {22.80} \\ 
                                        \cline{2-8}
                                        & \multirow{3}*{E5-large-v2}            &       -        & {70.18} & {72.50} & {76.73} & {\color{red}40.65} & {\color{red}18.03} \\ 
                                        \cline{3-8}
                                        &                                       &  GPT-4o & {\color{blue}72.69}   & {71.46}  & {75.87}  &  {\color{blue}50.43} & {20.50} \\  
                                        \cline{3-8}
                                        &                                       &  GPT-3.5-turbo & {72.23} & {71.88} & {\color{blue}78.29} & {50.16} & {\color{blue}23.08} \\ 
                                        \cline{3-8}
                                        &                                       &  Mistral-7b-instruct    & {69.92}  & {\color{blue}72.97}  & {76.90}  & {48.67} & {22.52} \\ 
                                        \cline{2-8}
                                        & \multirow{4}*{nomic-embed-text-v1.5}  &       -        & {\color{red}66.68} & {\color{red}67.28} & {79.37} & {\color{red}45.80} & {\color{red}23.93} \\ 
                                        \cline{3-8}
                                        &                                       &  GPT-4o & {\color{blue}71.45}   & {69.69}  & {78.60}  &  {45.94} & {24.22} \\  
                                        \cline{3-8}
                                        &                                       &  GPT-3.5-turbo & {68.87} & {67.80} & {\color{blue}80.42} & {\color{blue}46.05} & {25.73} \\ 
                                        \cline{3-8}
                                        &                                       &  Mistral-7b-instruct    & {69.20}  & {\color{blue}70.56}  & {78.83}  & {45.93} & {\color{blue}27.18} \\ 
                                        \cline{2-8}
                                        & \multirow{3}*{text-embedding-3-large} &       -        & {\color{red}72.52} & { 72.86} & { 83.81} & {54.14} & {26.25} \\  
                                        \cline{3-8}
                                        &                                       &  GPT-4o & {\color{blue}75.57}   & {72.24}  & {83.40}  &  {54.33} & {25.49} \\  
                                        \cline{3-8}
                                        &                                       &  GPT-3.5-turbo & {74.44} & {72.18} & {\color{blue}83.59} & {53.85} & {\color{blue}27.36} \\ 
                                        \cline{3-8}
                                        &                                       &  Mistral-7b-instruct    & {73.97}  & {72.44}  & {83.30}  & {\color{blue}54.51} & {26.99} \\ 
\bottomrule
\end{tabular}
}
\caption{
NDCG@10 results produced by different retrievers, embedding models, and hypothetical query generators (LLMs) across various datasets. The `$-$’ sign indicates that the results in the associated row are generated with the baseline embedding model. The red color indicates that the baseline embedding model is outperformed by HyQE with all three LLMs. 
The blue color indicates that the highest NDCG@10 value for a combination of retriever and embedding models under a dataset is achieved by HyQE. 
According to the MTEB leaderboard \cite{mteb}, increasing NDCG@10 by $1$ can improve the ranking by up to $10$ positions.
}
\label{tab:main_result}
\end{table*}
We test our method on multiple benchmarks to investigate the main question: {\it whether HyQE improves the nDCG@10 in the benchmarks?} 
In addition, we also investigate the following questions.
\begin{enumerate}[label=\Alph*., align=left, leftmargin=*, nosep]
    \item  Does changing the LLMs influence the results?
    \item  How many hypothetical queries does an LLM need to generate for each context?
    \item  Does changing the $\lambda$ in Eq.\ref{eq:hyqe} influence the results?
    \item  Is HyQE compatible with different retrieval methods such as HyDE \cite{hyde}?
    \item  How well does Eq.\ref{eq:hyqe_alt} perform in comparison with Eq.\ref{eq:hyqe}?
\end{enumerate}

\noindent{\bf Datasets.}  We test our methods on the following datasets: COVID~\cite{beir}, NEWS~\cite{beir}, Touche2020~\cite{beir}, DL19~\cite{trec}, and DL20~\cite{trec}. 
We use the same prompt for all the datasets except for the touche2020 dataset, in which the queries represent topics of arguments while the contexts consist of dialogues in those arguments. 
The prompt designed for this dataset can be found in Appendix \ref{sec:app_1}.

\noindent{\bf Baselines.} 
We use two kinds of retrievers:
one is embedding model-based retrievers, including contriever and bge-base-en-v1.5; the other is SPLADE++\_EnsembleDistil \cite{splade}, which is a sparse retrieval model that does not generate text embeddings.
We use the pre-built Lucene indexes in Pyserini \cite{pyserini} for retrieval.
We use five embedding models as the baselines for ranking: contriever~\cite{contriever}, bge-base-en-v1.5 \cite{bge}, E5-large-v2 \cite{e5}, text-embedding-3-large, and nomic-embed-text-v1.5 \cite{nomic}. 
We also use those embedding models as the backbones of HyQE and compare the results produced by HyQE with those produced by the baseline embedding models.
We use three different LLMs to generate the hypothetical queries: Mistral-7b-instruct-v0.2 \cite{mistral7b}, GPT-3.5 turbo, and GPT-4o.

\noindent{\bf Implementation Details.} We first retrieve $100$ contexts with a retriever.
Then, we use an embedding model to rank the contexts based on the cosine similarity between the context and the query and produce an ordered-set $C_{q, K}$ of candidate contexts where we set $K=30$. 
Then, we use the proposed method to obtain $r_q$ and re-rank these $30$ contexts.
Then, we compare these results with the ranking produced by the embedding model.

\begin{figure*}[!ht]
     \centering
        \begin{subfigure}[b]{0.24\textwidth}
         \centering
         \includegraphics[height=4.5cm, width=4.7cm]{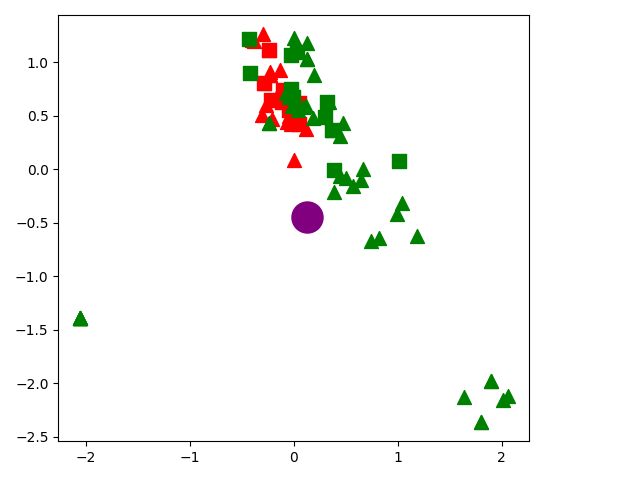}
        \end{subfigure}
        \begin{subfigure}[b]{0.24\textwidth}
         \centering
         \includegraphics[height=4.5cm, width=4.7cm]{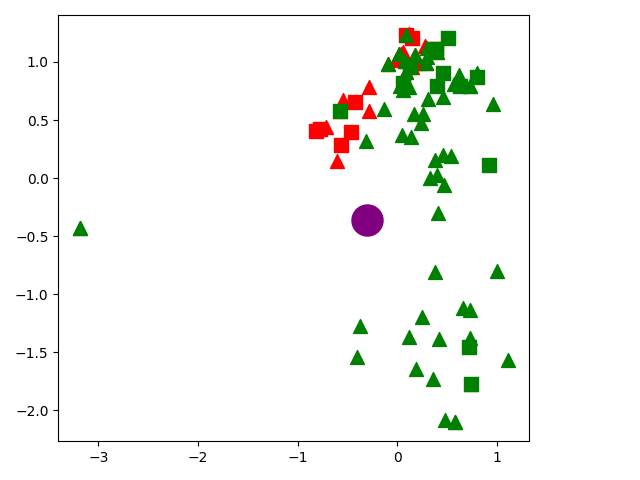}
        \end{subfigure}
        \begin{subfigure}[b]{0.24\textwidth}
         \centering
         \includegraphics[height=4.5cm, width=4.7cm]{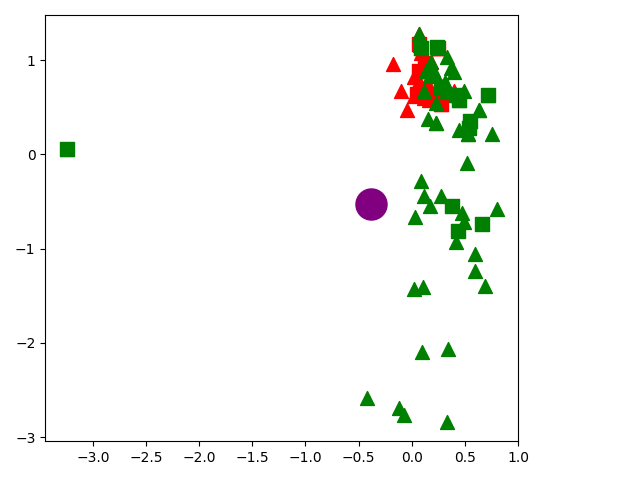}
        \end{subfigure}
        \begin{subfigure}[b]{0.24\textwidth}
         \centering
         \includegraphics[height=4.5cm, width=4.7cm]{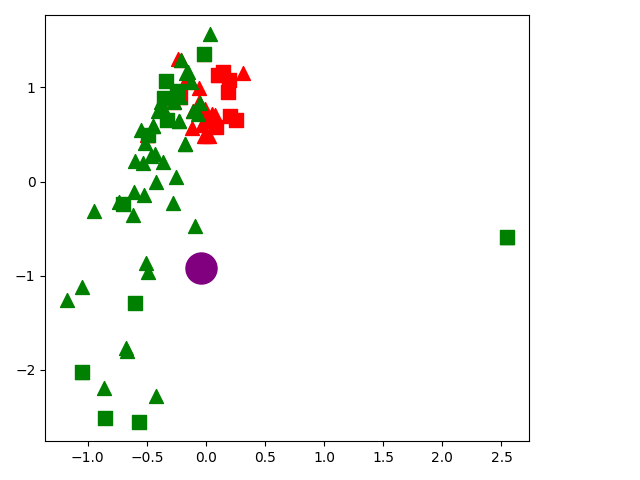}
        \end{subfigure}
 \caption{
 ICA on the bge-base-env-v1.5 embeddings for $4$ queries from COVID dataset. 
 Each figure corresponds to one query.
 The large purple circle represent a query in the dataset. 
 The red squares represent the top 5 contexts ranked using cosine similarity, and the red triangles represent the corresponding hypothetical queries.
 The green squares represent the top 5 contexts ranked using our method, and the green triangles represent the corresponding hypothetical queries. 
 The full analysis on all the $50$ queries can be found in Appendix \ref{sec:app_0}.
 } 
 \label{fig:ica_preview}
\end{figure*}

\noindent{\bf Main Results.} 
Table \ref{tab:main_result} shows the NDCG@10 produced by our methods and baseline embedding models on the benchmarks.
The Retrieval Model and Embedding Model columns indicate which models provide the initial list of $100$ contexts and which model is used for providing the $C_{q,30}$ candidate contexts.
The HyQE Model column indicates which LLMs are used to generate the hypothetical queries. 
The symbol `$-$' indicates that the results in the associated rows are produced by the baseline embedding models without hypothetical queries.
The other rows are obtained by HyQE framework with different combinations of retrieval models, embedding models, and hypothetical query generators.
Our methods outperform the associated baseline embedding models most of the time.
These results answer our main question and Question A, showing that locally hosted small-sized models and closed-source proprietary large models can generate high-quality hypothetical queries that result in high-quality rankings in our framework.

In addition, we use independent component analysis (ICA) to visualize the difference between the contexts ranked by cosine similarity and those by HyQE. 
By projecting the high-rank contexts' embeddings onto a 2D plane, Fig.\ref{fig:ica_preview} shows that the contexts ranked by cosine similarity tend to cluster near the input query in the embedding space. 
In contrast, the contexts ranked by HyQE and their corresponding hypothetical queries are more scattered.
This suggests that, in the embedding space, the queries are not necessarily adjacent to the contexts that provide answers to them. 
Hence, both the ICA visualization and the main results support our proposition that cosine similarity should be applied only when comparing queries with queries to ensure better preservation of the causal structure and to avoid spurious correlations.

\begin{figure*}[!tp]
     \centering
    \begin{subfigure}[b]{0.32\textwidth}
         \centering
         \includegraphics[width=\textwidth]{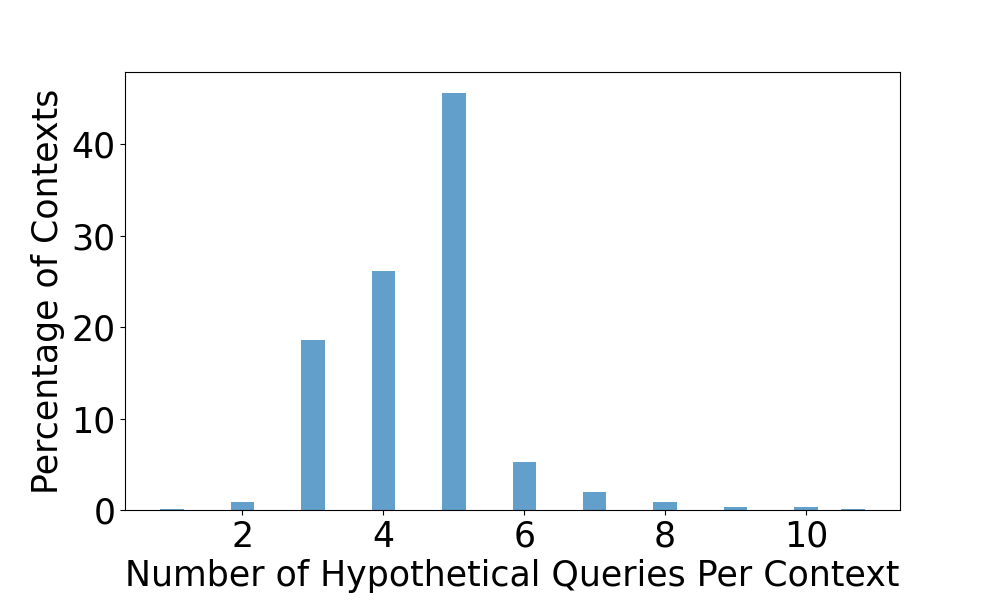}
         \caption{Gpt-3.5-Turbo on COVID}
     \end{subfigure} 
    \begin{subfigure}[b]{0.32\textwidth}
         \centering
         \includegraphics[width=\textwidth]{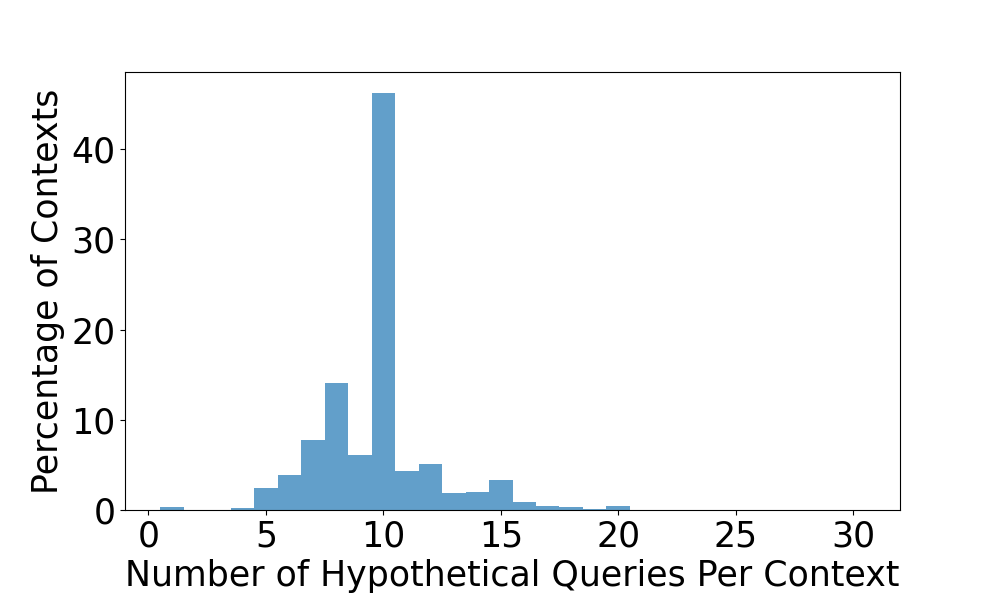}
         \caption{Gpt-4o on COVID}
     \end{subfigure} 
     \begin{subfigure}[b]{0.32\textwidth}
         \centering
         \includegraphics[width=\textwidth]{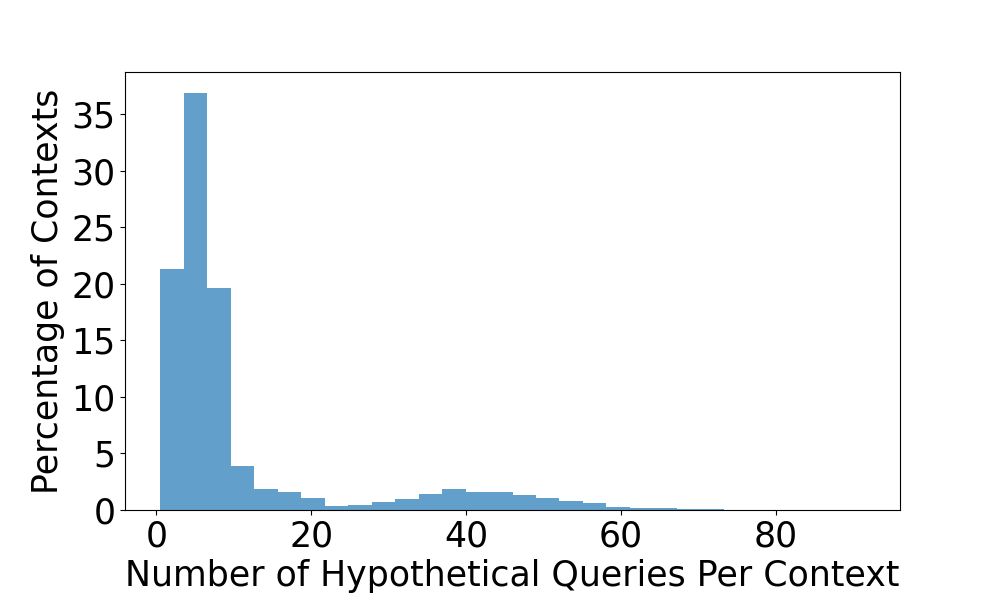}
         \caption{Mistral-7b-instruct on COVID}
     \end{subfigure} 
     
    \caption{
    The statistics of the number of hypothetical queries generated for the contexts in COVID datasets.
    The x-axis indicates the number of hypothetical queries generated for a context. 
    The y-axis indicates the percentage of contexts in the dataset. 
    The full results on all the dataset can be found in Appendix.\ref{sec:app_2}.    
    }
    \label{fig:num_queries_preview}
\end{figure*}

\begin{table}
\centering
\resizebox{0.99\columnwidth}{!}{
\begin{tabular}{m{7.5em} | m{5em} | m{2.3em}  m{2.3em}  m{2.3em}   m{2.3em}   m{2.3em}} 
\toprule
Embedding Model & Downsample & DL19 & DL20 & COVID & NEWS & Touche \\
\hline
\multirow{2}*{contriever}             &  $10\%$          &	45.90 &	42.47 & 33.04 &	37.94 &	16.94 \\  
                                      & $50\%$ &	51.49 &	49.68&	34.65 &	40.23 &	16.71\\  
\hline
\multirow{2}*{bge-base-en-v1.5}      &  $10\%$  &	68.21 &	64.66 &	77.69 &	42.44 &	19.56  \\  
                                      & $50\%$ &	70.05 & 67.24 & 79.38 & 44.24 & 21.36\\                      
\hline 
\bottomrule                               
\end{tabular}
 }
\caption{ Average NDCG@10 after randomly downsampling the hypothetical queries generated by GPT-4o by different ratios for multiple times. 
}
\label{tab:downsample}
\end{table}  

\noindent{\bf Changing the number of hypothesis queries.} 
As indicated by the prompt in Figure \ref{fig:prompt}, the number of the generated hypothesis queries for each context is determined by the LLM.
Our statistics in Fig.\ref{fig:num_queries_preview} shows that the LLMs generate less then $20$ hypothetical queries for most of the contexts. 
To verify if reducing the number of hypothetical queries can influence the performance, we downsample the hypothetical queries generated by an LLM at a determined ratio multiple times, and then averaged the results.
Table \ref{tab:downsample} shows that after downsampling the queries, most of the NDCG@10 scores are slightly lower than those without downsampling in Tabel \ref{tab:main_result}.
We hypothesize that it is because different hypothetical queries can be far from each other in the embedding space, which can be observed in Fig.\ref{fig:ica_preview}.
Since in Eq.\ref{eq:hyqe} we use $\max$ to identify the most relevant hypothetical queries via cosine-similarity, downsampling the queries by half, as in our case, can easily exclude relevant hypothetical queries, and thus reduce the ranking performance.
However, even when downsampling only $10\%$ of the queries, most of the results are still better than those without HyQE as shown in Table \ref{tab:main_result}.
\begin{figure}[tp!]
    \begin{subfigure}[t]{0.8\textwidth}
         \includegraphics[height=0.36cm, width=8cm]{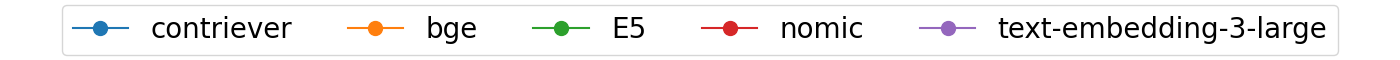}
    \end{subfigure} 
    \vspace{-5mm}
    %
    \\
    \begin{subfigure}[b]{0.24\textwidth}
         \centering
         \includegraphics[height=3.2cm, width=4.25cm]{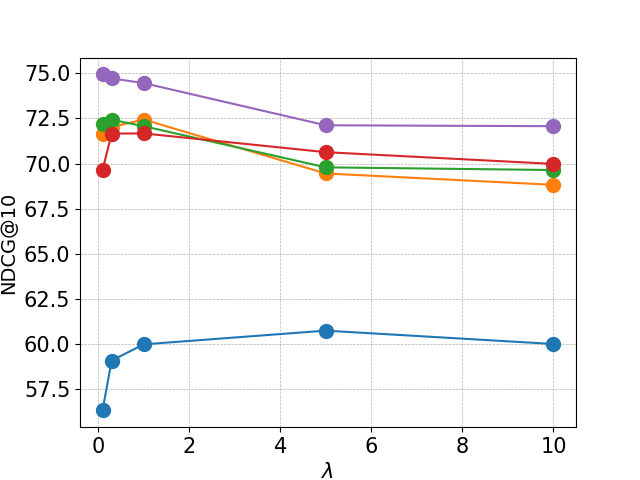}
         \caption{DL19}
     \end{subfigure}%
     \hfill
     \begin{subfigure}[b]{0.24\textwidth}
         \centering
         \includegraphics[height=3.2cm, width=4.25cm]{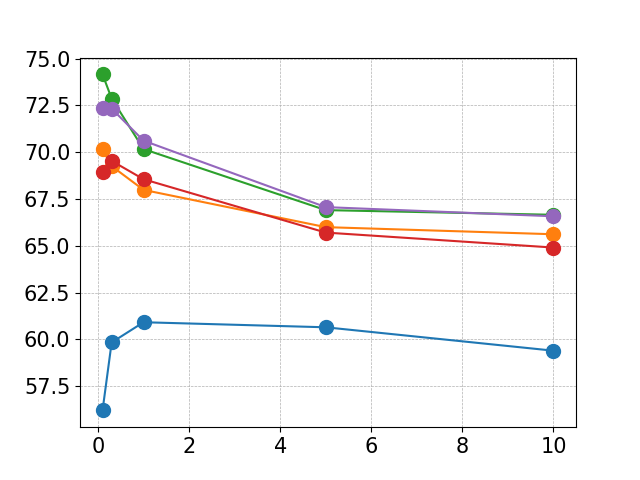}
         \caption{DL20}
     \end{subfigure}
        \caption{NDCG@10 changes with $\lambda$.
        } 
    \label{fig:par}
\end{figure}

\noindent{\bf Changing the hyperparameter $\lambda$.} Next, we answer Question C by changing the hyperparameters $\lambda$ in Eq.\ref{eq:hyqe} to examine how sensitive the HyQE framework is to the changes.
We pick $2$ datasets, $4$ embedding models, i.e., contriever, bge-base-en-v1.5, E5-large-v2, and nomic-embed-text-v1.5, and use SPLADE++ ED as the retriever so that the candidate contexts are the same. 
Fig.\ref{fig:par} shows that NDCG@10 decreases as $\lambda$ increases for most embedding models, suggesting choosing small $\lambda$ for these models.
In Appendix \ref{sec:app_2}, we will present the results of modifying $\lambda$ for other datasets, and we will also explore the impact of changing the number of candidate contexts, i.e., the $K$ in $C_{q,K}$, from 30 to other values.

\noindent{\bf Compatibility with HyDE.} To examine whether HyQE is compatible with other methods, we combine our method with HyDE \cite{hyde} by using HyDE for context retrieval and HyQE for context ranking.
We use the identical embedding models for context retrieval and ranking, and use GPT-4o for the hypothetical context and query generation.
Since HyDE generates hypothetical contexts and uses the average of the query embedding and hypothetical embeddings for context retrieval, we implement this combination in two ways.
The first is to only use HyDE to collect $100$ contexts and repeat the context ranking with HyQE as in Algorithm \ref{algo:hyqe}.
The second is to use HyDE to not only collect the $100$ contexts but also replace the query embedding with the mean of the query and hypothetical context embeddings during execution of Algorithm \ref{algo:hyqe}. 
In Table \ref{tab:hyde}, we compare the results obtained in these two ways as well as those of using HyDE alone.
The results answer Question C by showing that HyQE is not only compatible with HyDE but also can further improve the ranking quality beyond that in Table \ref{tab:main_result}.  
\begin{table}
\centering
\resizebox{0.99\columnwidth}{!}{
\begin{tabular}{m{8em} | m{3em} | m{2.3em}  m{2.3em}  m{2.3em}   m{2.3em}   m{2.3em}} 
\toprule
Embedding Model & HyDE & DL19 & DL20 & COVID & NEWS & Touche \\
\hline
\multirow{3}*{contriever}             &   -           & {\color{red}62.60}   & {\color{red}57.69}  & {\color{red}53.86}  & {\color{red}38.76} & {\color{red}17.92} \\  
                                      \cline{2-7}
                                      &    $+$HyQE    & {65.58}   & {62.72}  & {54.39}  & {43.59} & {18.81} \\  
                                      \cline{2-7}
                                      & $\times$HyQE  & {\color{blue}67.38}   & {\color{blue}63.35}  & {\color{blue}57.52}  & {\color{blue}45.49} & {\color{blue}20.41} \\  
                                          
\hline
 \multirow{3}*{bge-base-en-v1.5}      &  -               & {75.37}   & {\color{red}70.55}  & {\color{red}75.49}  & {\color{red}43.55} & {\color{red}17.92} \\  
                                      \cline{2-7}
                                      &    $+$HyQE       & {75.16}   & {71.36}  & {\color{blue}78.98}  & {46.12} & {\color{blue}20.69} \\  
                                      \cline{2-7}
                                      &    $\times$HyQE  & {\color{blue}75.96}   & {\color{blue}72.07}  & {78.81}  & {\color{blue}46.85} & {20.39} \\ 
\bottomrule                               
\end{tabular}
 }
\caption{ NDCG@10 results produced by combining HyDE with HyQE. 
In the `HyDE' column, the `-' symbol indicates that the results in the associated rows are generated by HyDE; 
`$\times$HyQE' indicates that after HyDE is used to retrieve contexts, the query embedding has been changed into the average embedding of the query and hypothetical contexts when HyQE ranks the contexts.
`$+$ HyQE' indicates that the query embedding is not changed when HyQE ranks the contexts;
The font color scheme is similar to that in Table \ref{tab:main_result}.
}
\label{tab:hyde}
\end{table}

\noindent{\bf Using the Alternative Scoring Function.} We next answer Question E by evaluating the alternative scoring function in Eq.\ref{eq:hyqe_alt}. 
We use two embedding models, i.e., contriever and bge-base-en-v1.5, for both context retrieval and ranking.
The hyperparameter $\lambda$ for each embedding model stays the same as that produces the main results.
We still use GPT-4o to generate hypothetical queries.
The results are included in Table \ref{tab:alt_hyqe}. 
By comparing with the results in Table \ref{tab:main_result}, it is obvious that using Eq.\ref{eq:hyqe_alt} outperforms the baseline embedding models and cannot outperform using Eq.\ref{eq:hyqe}.
\begin{table}
\centering
\resizebox{0.99\columnwidth}{!}
{
\begin{tabular}{m{8em} | m{2.3em}  m{2.3em}   m{2.3em}  m{2.3em}  m{2.3em}} 
\toprule
Embedding Model &  DL19 & DL20 & COVID & NEWS & Touche \\
\hline
contriever             & {51.33}   & {46.76}  & {33.10}  & {38.87} & {15.33}\\
                                          
\hline
{bge-base-en-v1.5}      & {71.04}   & {66.48}  & {79.52}  & {43.57} & {18.40}\\
\bottomrule
\end{tabular}
 }
\caption{ NDCG@10 produced by using Eq.\ref{eq:hyqe_alt} for HyQE.
}
\label{tab:alt_hyqe}
\end{table}

\section{Conclusion}
In this paper, we introduce a novel framework for context ranking using hypothetical queries generated by LLMs. 
Our method is grounded in variational inference, aiming to preserve the causal relationship between queries and the contexts. 
The experimental results demonstrate that our approach not only outperforms baselines but also can be integrated seamlessly with existing techniques, allowing for iterative refinement and continuous improvement. 
Furthermore, our method can amortize the overhead in text generation with LLM as the input queries increase, offering a scalable and efficient solution for context retrieval and ranking.

\section*{Acknowledgements}

This work was supported by the Intuit University Collaboration Program, and in part by the NSF under grant CCF-2340776.
\section*{Limitations}

While our proposed framework demonstrates significant improvements in context ranking and is scalable, there are several limitations to consider:
\begin{enumerate}
	\item	{\bf Overhead of Query Generation and Storage.} The effectiveness of our method relies on using an LLM to generate the queries.
    The computational complexity for the query generation is amortized as the input queries grow. 
    However, this amortization is built on the premise that the generated queries are stored for future retrieval. 
    And such storage will raise the memory complexity of this framework. 
    As a result, extremely large datasets could still pose challenges.
    
	\item	{\bf Dependency on the Type of Query.} The input query can have different types, e.g., questions asking for specific information, a sequence of keywords, etc.
    However, in the prompt we only ask the LLM to generate the questions that can be addressed by the context, which may not have different structures than the input query.
  
	\item	{\bf Adaptability to Context Chunk Sizes.} 
 Our framework has been validated on well-known TREC and MS-MARCO datasets, where the contexts are provided. 
 However, when dealing with document retrieval, the contexts are created by segmenting the documents into chunks.
 The documents may be segmented with different chunk sizes depending on the requirement.
 Each time the document is segmented, the hypothetical queries have to be regenerated from the contexts. 
This issue could potentially be mitigated by generating hypothetical queries from smaller, fixed-sized chunks of contexts and composing those queries for larger chunks of contexts.
 However, the specifics of this approach require further investigation to ensure its effectiveness and efficiency.
\end{enumerate}

Addressing these limitations in future work will be essential for enhancing the robustness, efficiency, and applicability of our proposed context ranking framework across a broader range of scenarios.

\bibliography{emnlp}

\clearpage
\appendix
\onecolumn

\section{Visualizing the Hypothetical Query Emebeddings}\label{sec:app_0}

We demonstrate the difference between the contexts ranked by cosine similarity and those by HyQE. 
We conduct an independent component analysis (ICA) on each high-dimensional text embedding and project the embeddings onto a 2D plane.

\begin{figure*}[!ht]
     \centering
        \begin{subfigure}[b]{0.138\textwidth}
         \centering
         \includegraphics[height=2.2cm, width=2.5cm]{assets/plots/bge-base-en-v1.5_bge-base-en-v1.5_combined_covid_q1.png}
        \end{subfigure}
        \begin{subfigure}[b]{0.138\textwidth}
         \centering
         \includegraphics[height=2.2cm, width=2.5cm]{assets/plots/bge-base-en-v1.5_bge-base-en-v1.5_combined_covid_q2.png}
        \end{subfigure}
        \begin{subfigure}[b]{0.138\textwidth}
         \centering
         \includegraphics[height=2.2cm, width=2.5cm]{assets/plots/bge-base-en-v1.5_bge-base-en-v1.5_combined_covid_q3.png}
        \end{subfigure}
        \begin{subfigure}[b]{0.138\textwidth}
         \centering
         \includegraphics[height=2.2cm, width=2.5cm]{assets/plots/bge-base-en-v1.5_bge-base-en-v1.5_combined_covid_q4.png}
        \end{subfigure}
        \begin{subfigure}[b]{0.138\textwidth}
         \centering
         \includegraphics[height=2.2cm, width=2.5cm]{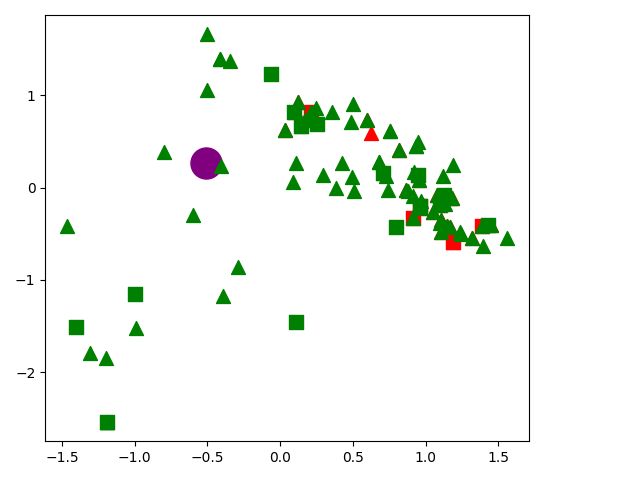}
        \end{subfigure}
        \begin{subfigure}[b]{0.138\textwidth}
         \centering
         \includegraphics[height=2.2cm, width=2.5cm]{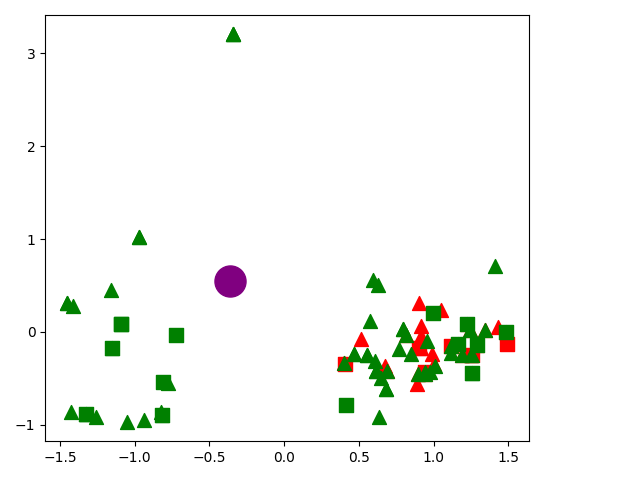}
        \end{subfigure}
        \begin{subfigure}[b]{0.138\textwidth}
         \centering
         \includegraphics[height=2.2cm, width=2.5cm]{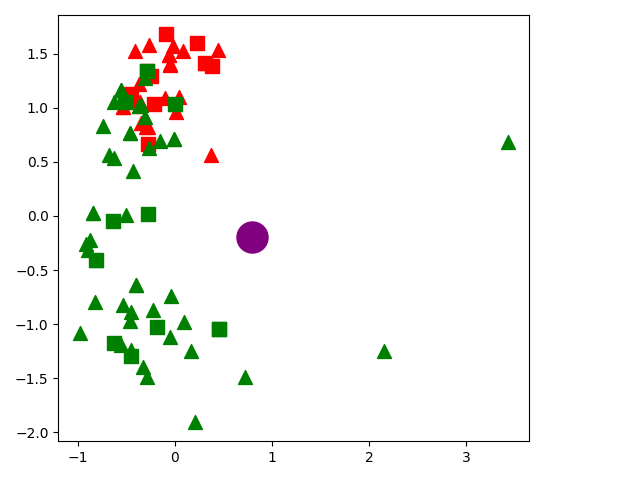}
        \end{subfigure}
        \begin{subfigure}[b]{0.138\textwidth}
         \centering
         \includegraphics[height=2.2cm, width=2.5cm]{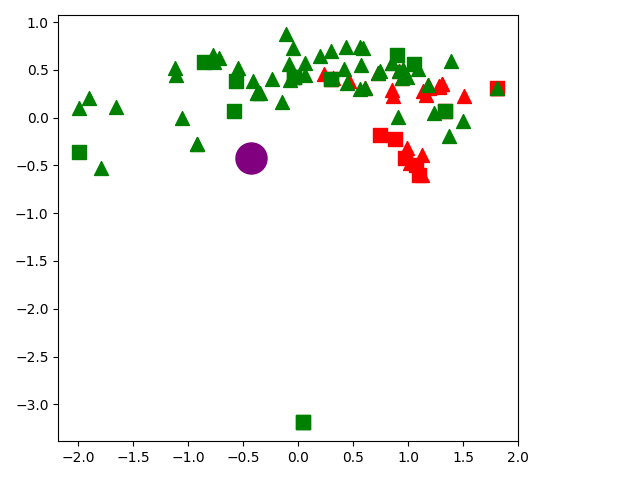}
        \end{subfigure}
        \begin{subfigure}[b]{0.138\textwidth}
         \centering
         \includegraphics[height=2.2cm, width=2.5cm]{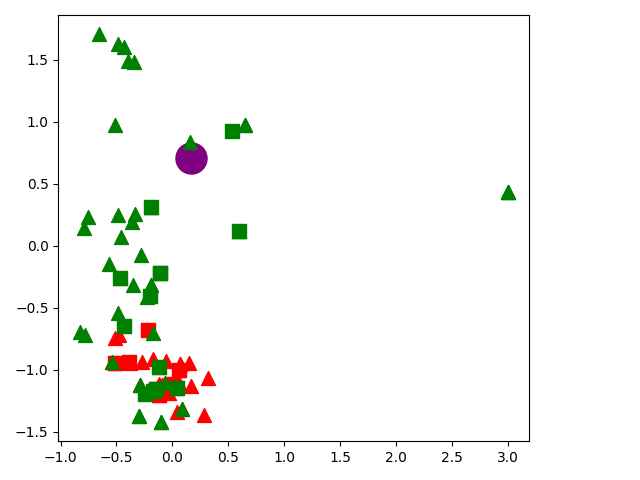}
        \end{subfigure}
        \begin{subfigure}[b]{0.138\textwidth}
         \centering
         \includegraphics[height=2.2cm, width=2.5cm]{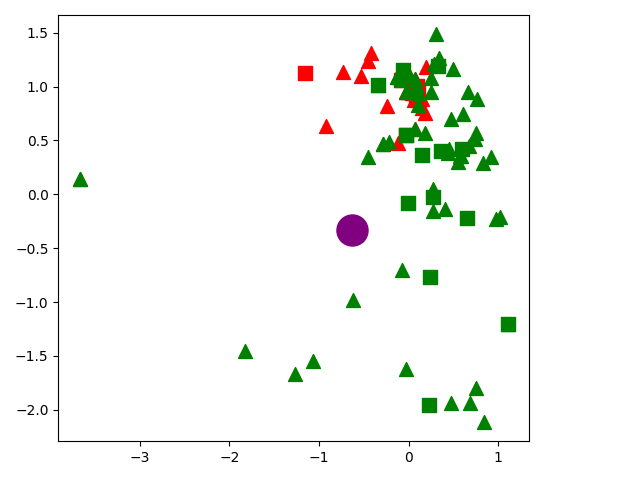}
        \end{subfigure}
        \begin{subfigure}[b]{0.138\textwidth}
         \centering
         \includegraphics[height=2.2cm, width=2.5cm]{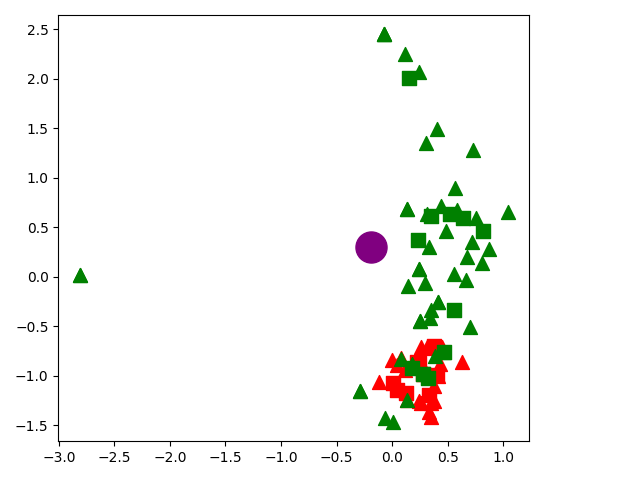}
        \end{subfigure}
        \begin{subfigure}[b]{0.138\textwidth}
         \centering
         \includegraphics[height=2.2cm, width=2.5cm]{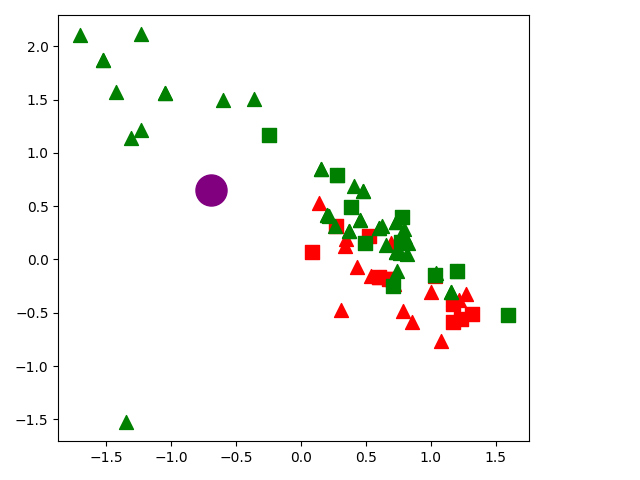}
        \end{subfigure}
        \begin{subfigure}[b]{0.138\textwidth}
         \centering
         \includegraphics[height=2.2cm, width=2.5cm]{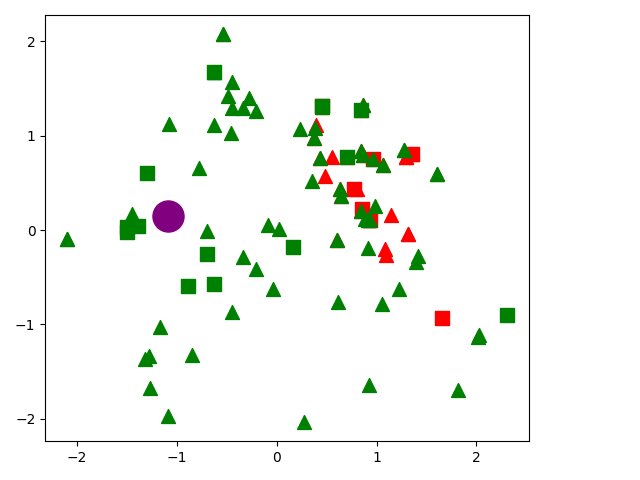}
        \end{subfigure}
        \begin{subfigure}[b]{0.138\textwidth}
         \centering
         \includegraphics[height=2.2cm, width=2.5cm]{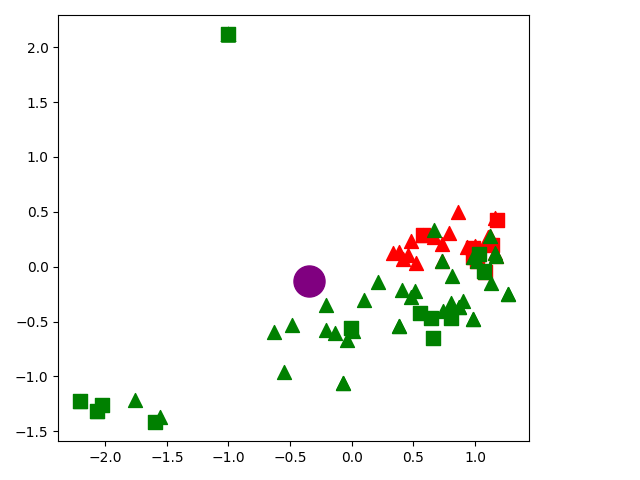}
        \end{subfigure}
        \begin{subfigure}[b]{0.138\textwidth}
         \centering
         \includegraphics[height=2.2cm, width=2.5cm]{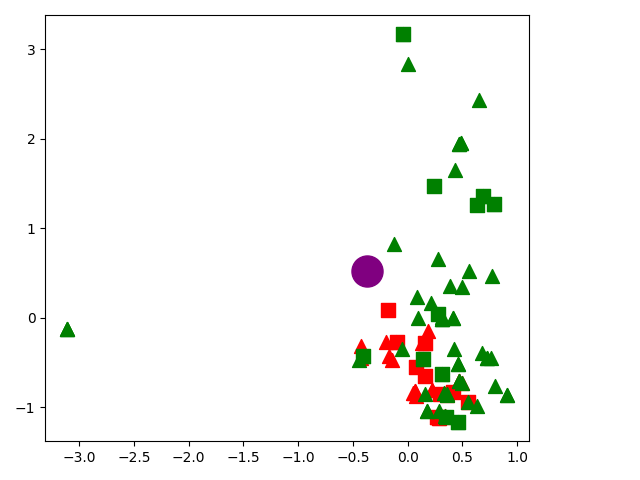}
        \end{subfigure}
        \begin{subfigure}[b]{0.138\textwidth}
         \centering
         \includegraphics[height=2.2cm, width=2.5cm]{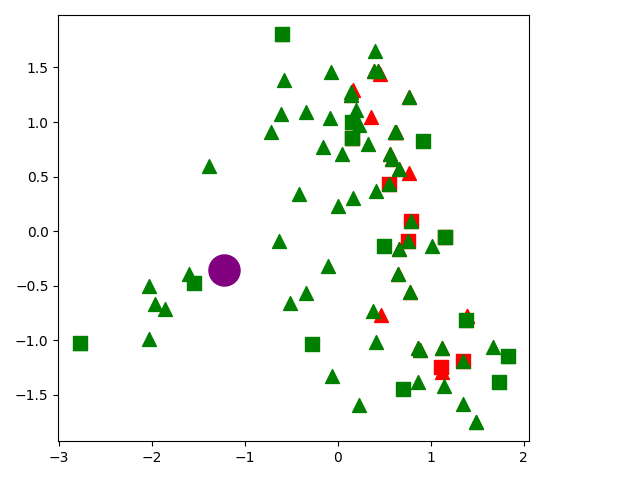}
        \end{subfigure}
        \begin{subfigure}[b]{0.138\textwidth}
         \centering
         \includegraphics[height=2.2cm, width=2.5cm]{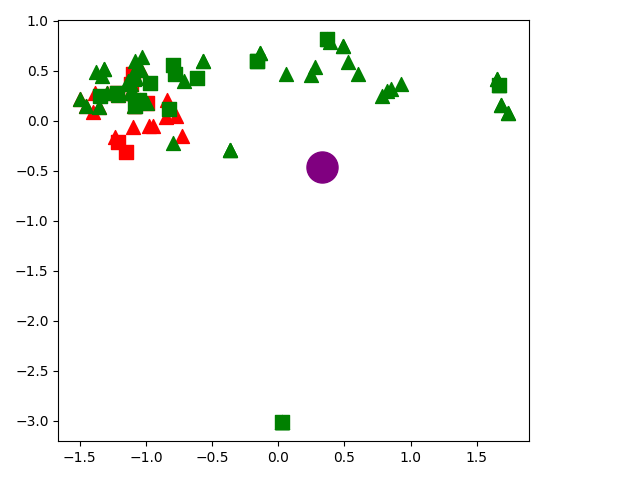}
        \end{subfigure}
        \begin{subfigure}[b]{0.138\textwidth}
         \centering
         \includegraphics[height=2.2cm, width=2.5cm]{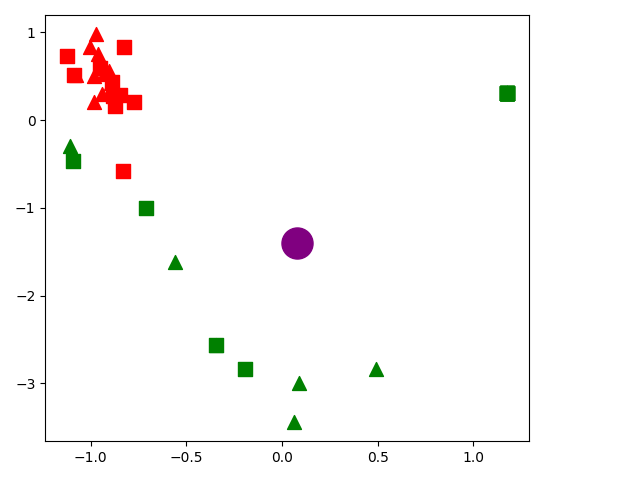}
        \end{subfigure}
        \begin{subfigure}[b]{0.138\textwidth}
         \centering
         \includegraphics[height=2.2cm, width=2.5cm]{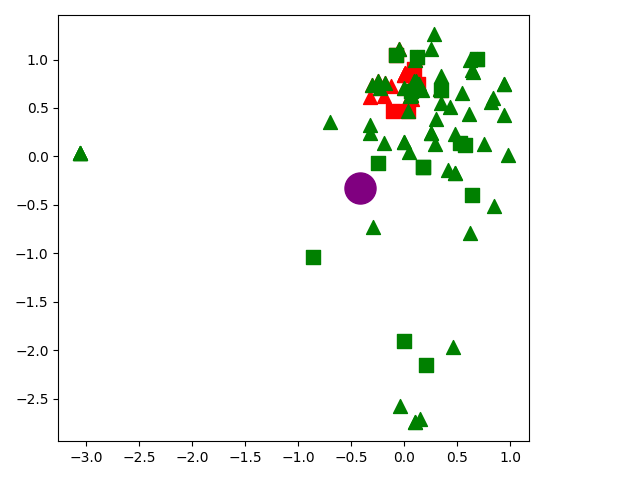}
        \end{subfigure}
        \begin{subfigure}[b]{0.138\textwidth}
         \centering
         \includegraphics[height=2.2cm, width=2.5cm]{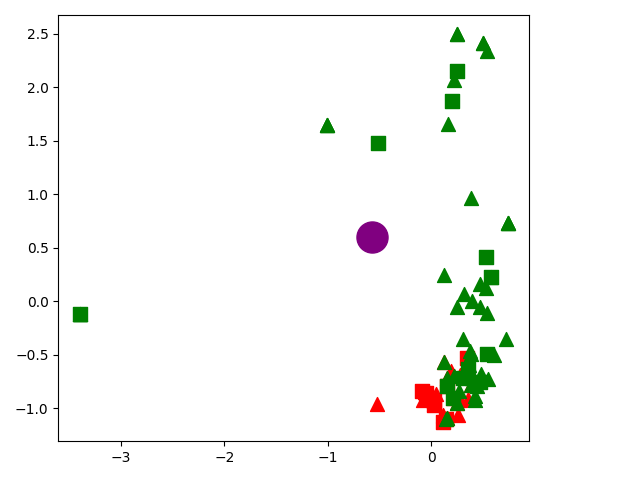}
        \end{subfigure}
        \begin{subfigure}[b]{0.138\textwidth}
         \centering
         \includegraphics[height=2.2cm, width=2.5cm]{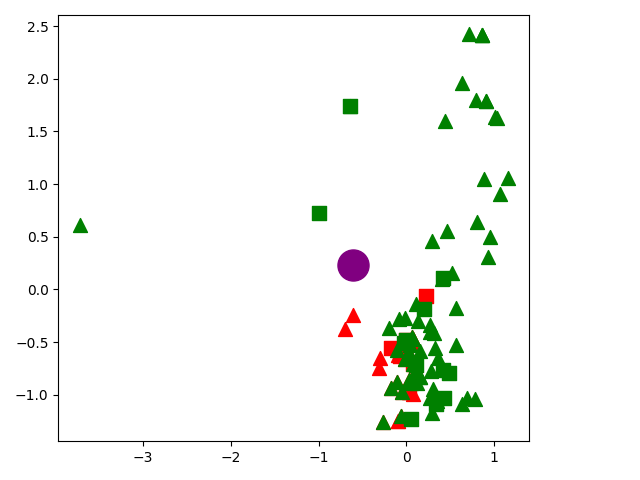}
        \end{subfigure}
        \begin{subfigure}[b]{0.138\textwidth}
         \centering
         \includegraphics[height=2.2cm, width=2.5cm]{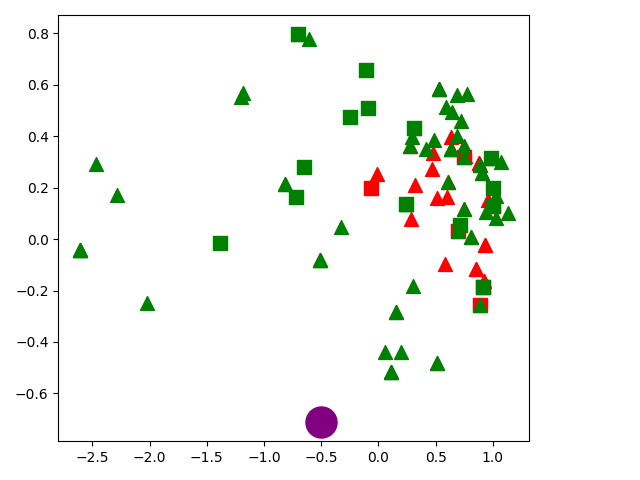}
        \end{subfigure}
        \begin{subfigure}[b]{0.138\textwidth}
         \centering
         \includegraphics[height=2.2cm, width=2.5cm]{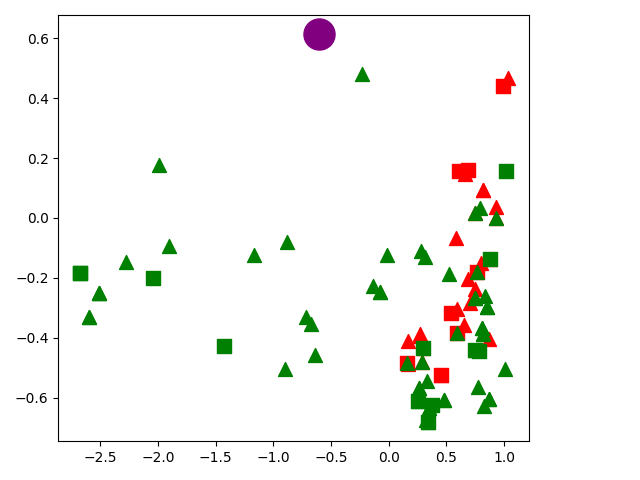}
        \end{subfigure}
        \begin{subfigure}[b]{0.138\textwidth}
         \centering
         \includegraphics[height=2.2cm, width=2.5cm]{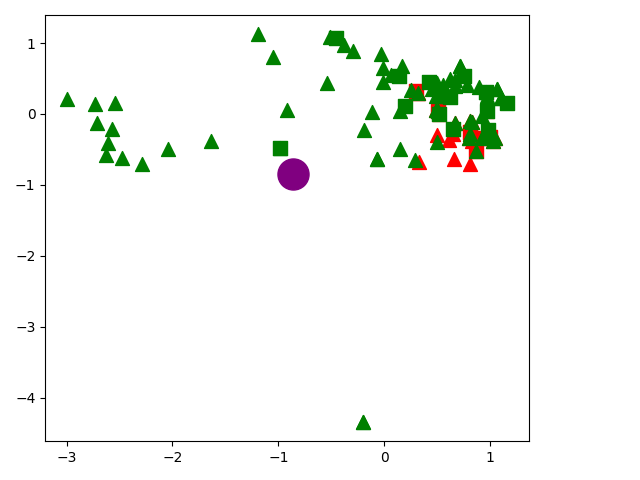}
        \end{subfigure}
        \begin{subfigure}[b]{0.138\textwidth}
         \centering
         \includegraphics[height=2.2cm, width=2.5cm]{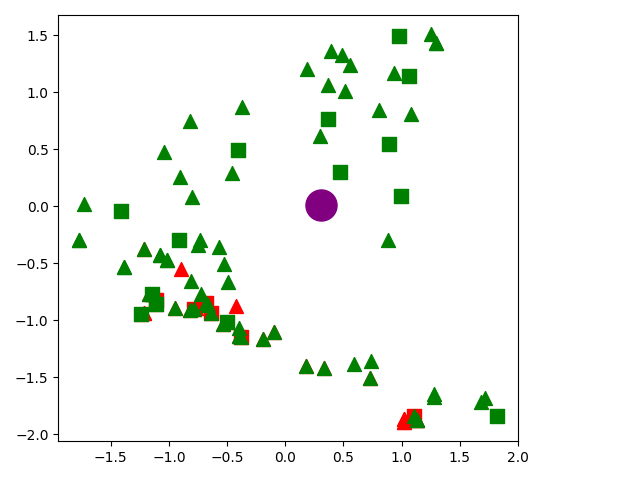}
        \end{subfigure}
        \begin{subfigure}[b]{0.138\textwidth}
         \centering
         \includegraphics[height=2.2cm, width=2.5cm]{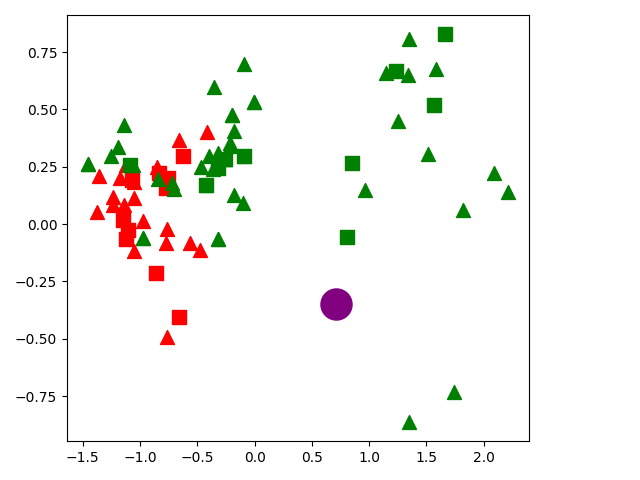}
        \end{subfigure}
        \begin{subfigure}[b]{0.138\textwidth}
         \centering
         \includegraphics[height=2.2cm, width=2.5cm]{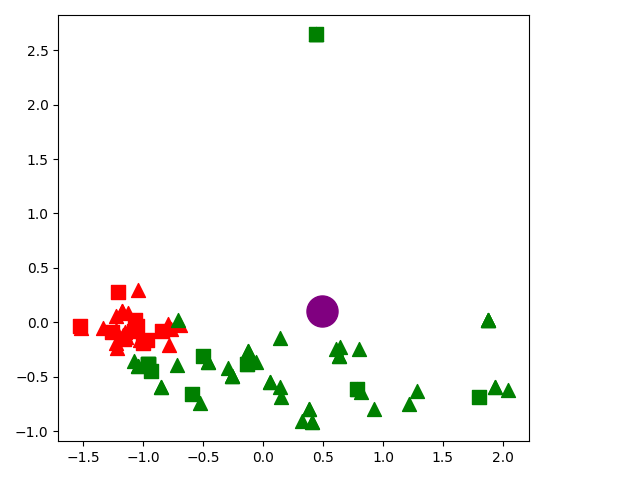}
        \end{subfigure}
        \begin{subfigure}[b]{0.138\textwidth}
         \centering
         \includegraphics[height=2.2cm, width=2.5cm]{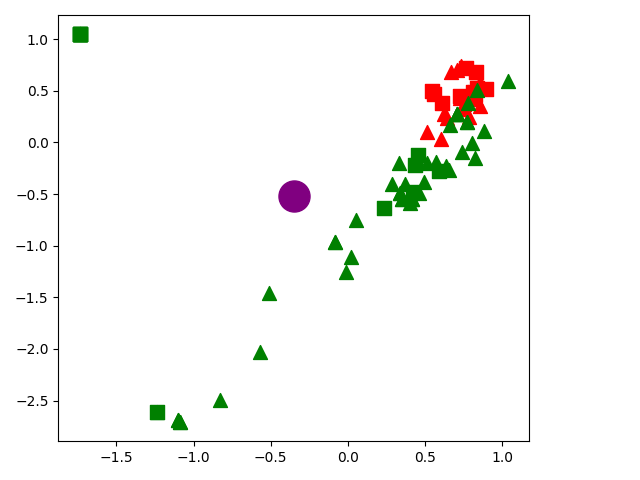}
        \end{subfigure}
        \begin{subfigure}[b]{0.138\textwidth}
         \centering
         \includegraphics[height=2.2cm, width=2.5cm]{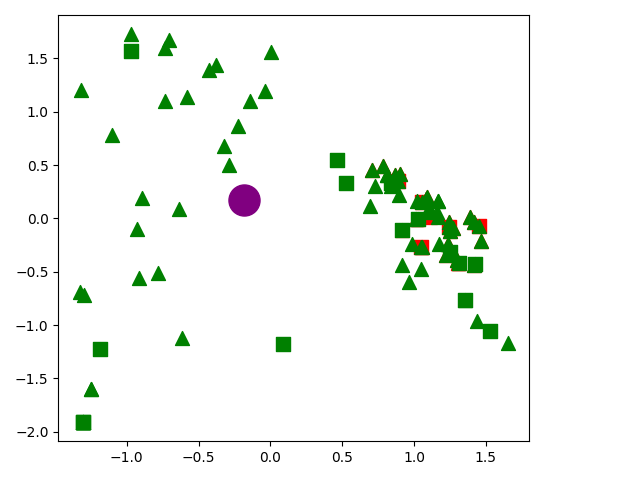}
        \end{subfigure}
        \begin{subfigure}[b]{0.138\textwidth}
         \centering
         \includegraphics[height=2.2cm, width=2.5cm]{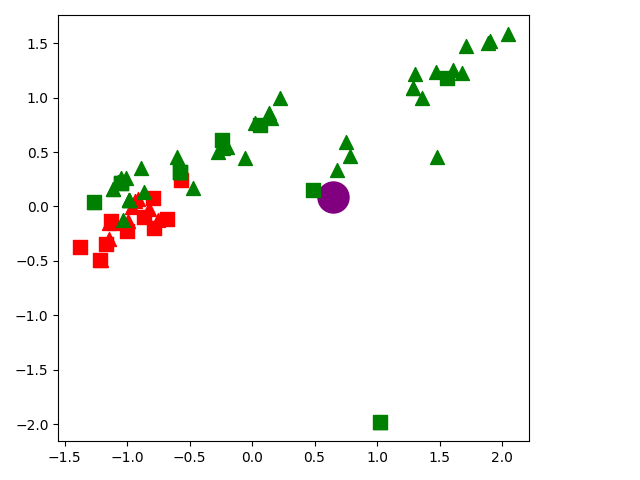}
        \end{subfigure}
        \begin{subfigure}[b]{0.138\textwidth}
         \centering
         \includegraphics[height=2.2cm, width=2.5cm]{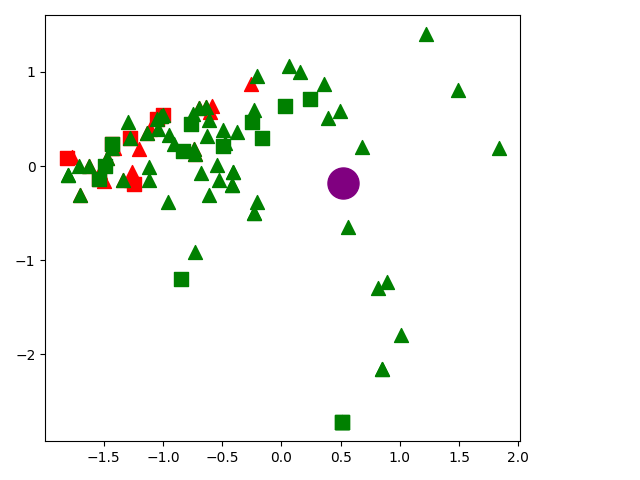}
        \end{subfigure}
        \begin{subfigure}[b]{0.138\textwidth}
         \centering
         \includegraphics[height=2.2cm, width=2.5cm]{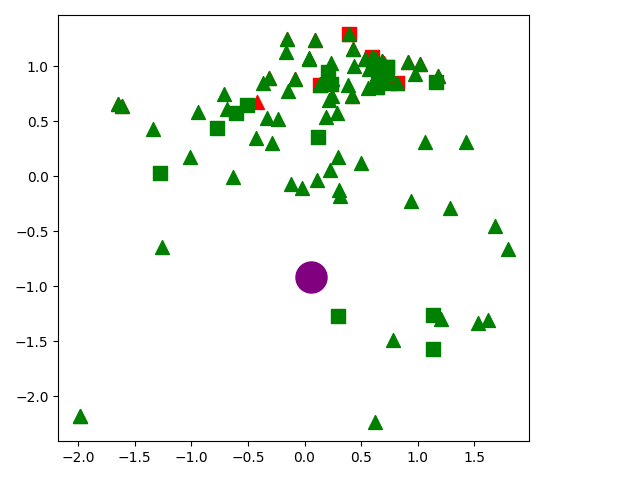}
        \end{subfigure}
        \begin{subfigure}[b]{0.138\textwidth}
         \centering
         \includegraphics[height=2.2cm, width=2.5cm]{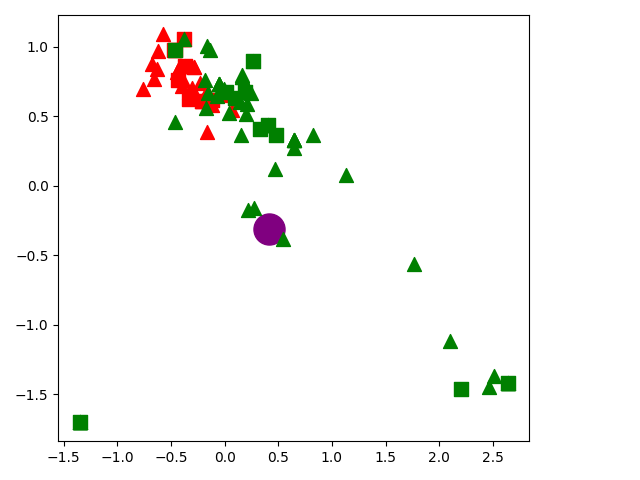}
        \end{subfigure}
        \begin{subfigure}[b]{0.138\textwidth}
         \centering
         \includegraphics[height=2.2cm, width=2.5cm]{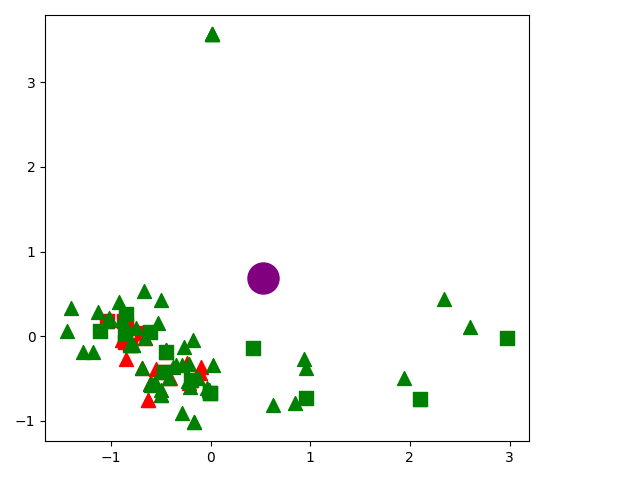}
        \end{subfigure}
        \begin{subfigure}[b]{0.138\textwidth}
         \centering
         \includegraphics[height=2.2cm, width=2.5cm]{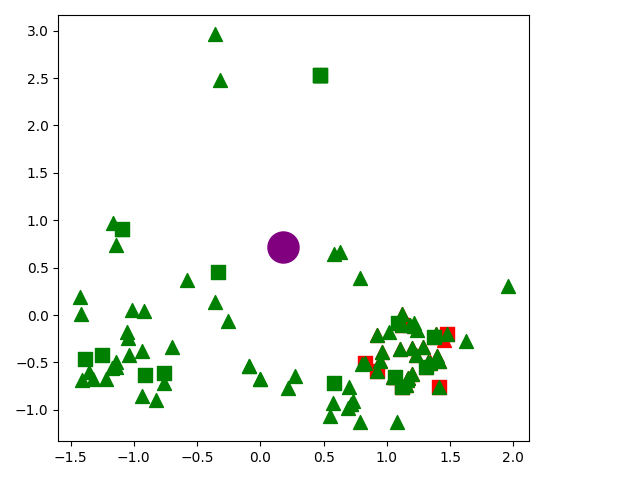}
        \end{subfigure}
        \begin{subfigure}[b]{0.138\textwidth}
         \centering
         \includegraphics[height=2.2cm, width=2.5cm]{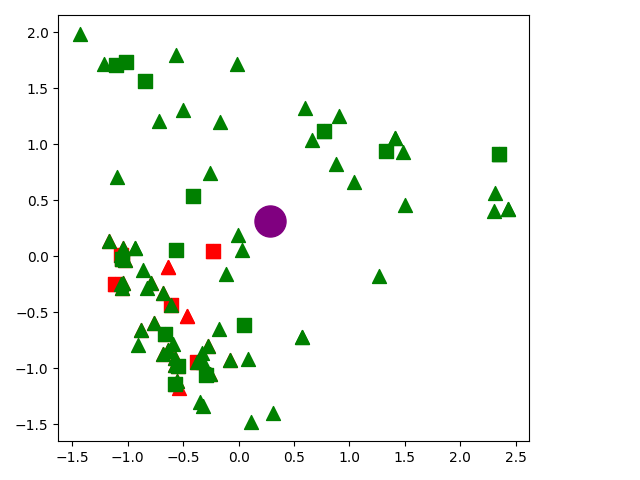}
        \end{subfigure}
        \begin{subfigure}[b]{0.138\textwidth}
         \centering
         \includegraphics[height=2.2cm, width=2.5cm]{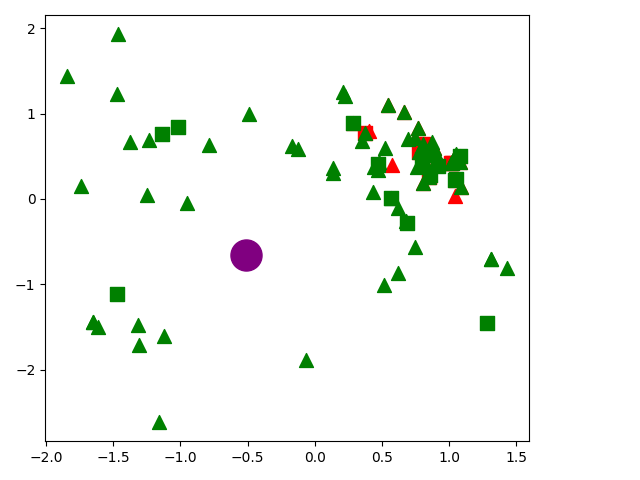}
        \end{subfigure}
        \begin{subfigure}[b]{0.138\textwidth}
         \centering
         \includegraphics[height=2.2cm, width=2.5cm]{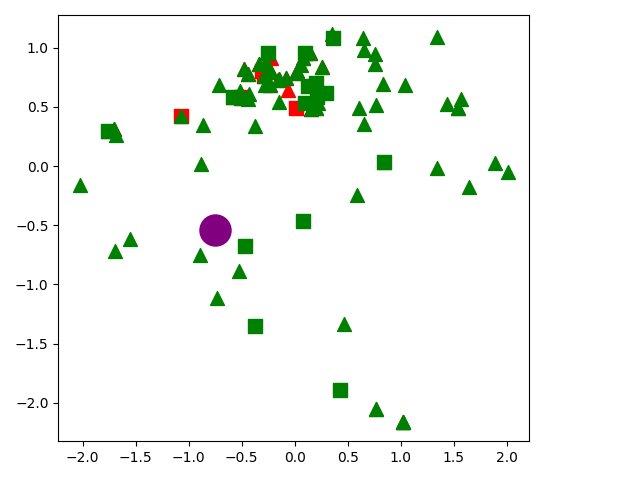}
        \end{subfigure}
        \begin{subfigure}[b]{0.138\textwidth}
         \centering
         \includegraphics[height=2.2cm, width=2.5cm]{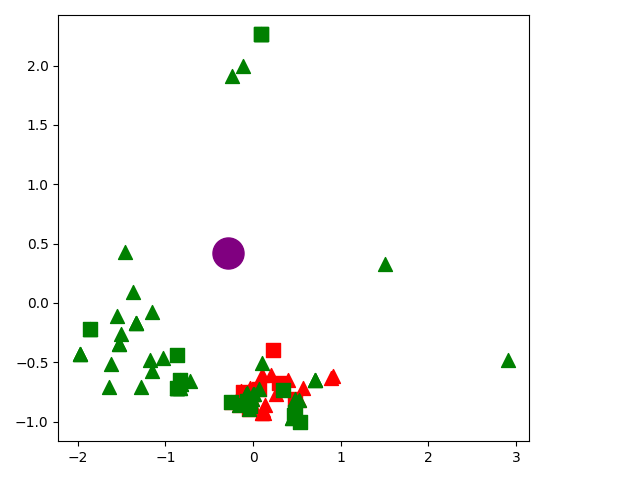}
        \end{subfigure}
        \begin{subfigure}[b]{0.138\textwidth}
         \centering
         \includegraphics[height=2.2cm, width=2.5cm]{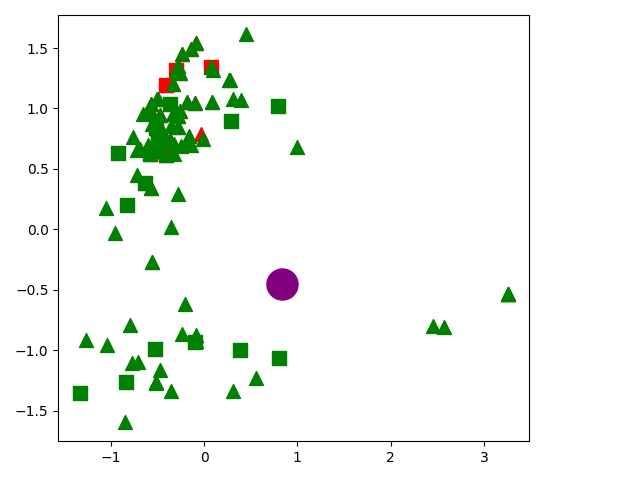}
        \end{subfigure}
        \begin{subfigure}[b]{0.138\textwidth}
         \centering
         \includegraphics[height=2.2cm, width=2.5cm]{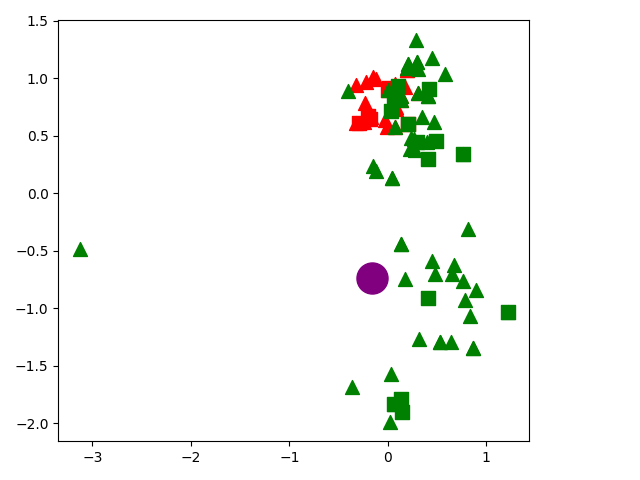}
        \end{subfigure}
        \begin{subfigure}[b]{0.138\textwidth}
         \centering
         \includegraphics[height=2.2cm, width=2.5cm]{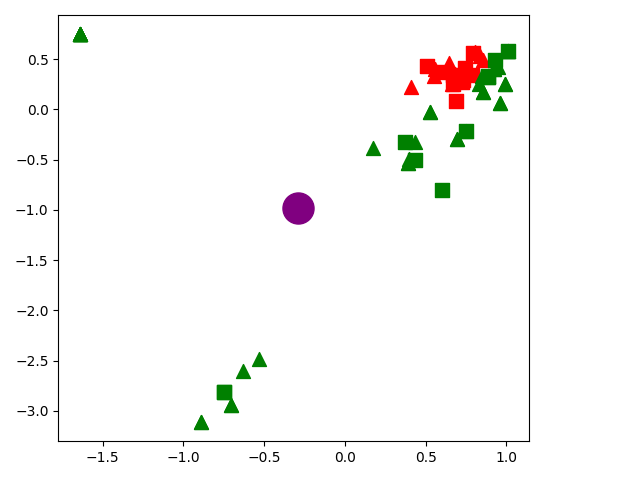}
        \end{subfigure}
         \begin{subfigure}[b]{0.138\textwidth}
         \centering
         \includegraphics[height=2.2cm, width=2.5cm]{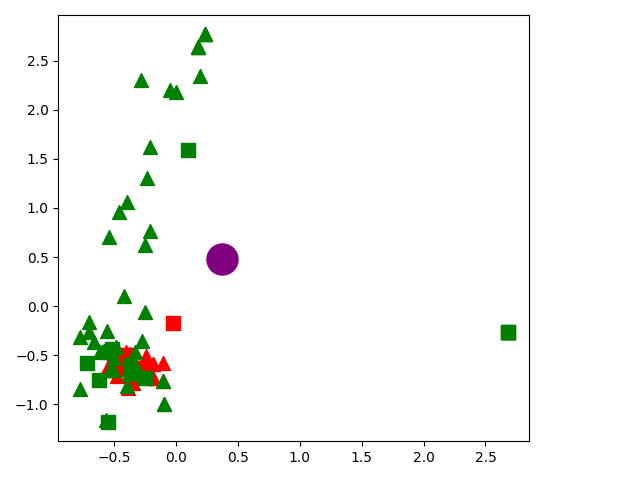}
        \end{subfigure}
         \begin{subfigure}[b]{0.138\textwidth}
         \centering
         \includegraphics[height=2.2cm, width=2.5cm]{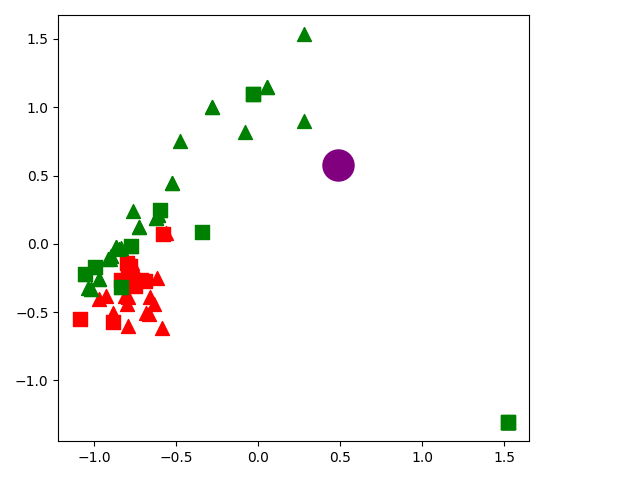}
        \end{subfigure}
         \begin{subfigure}[b]{0.138\textwidth}
         \centering
         \includegraphics[height=2.2cm, width=2.5cm]{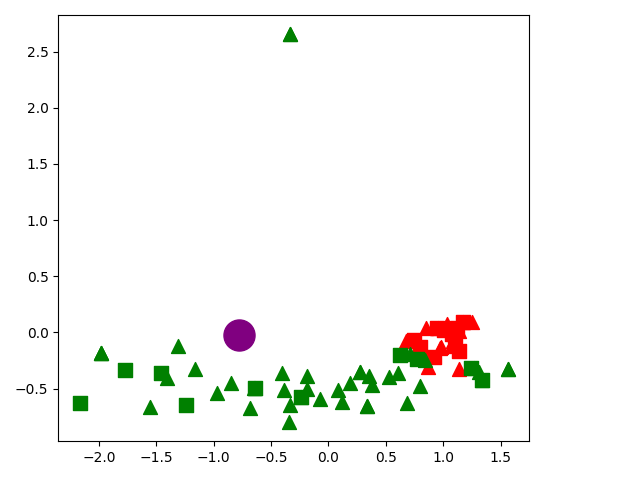}
        \end{subfigure}
         \begin{subfigure}[b]{0.138\textwidth}
         \centering
         \includegraphics[height=2.2cm, width=2.5cm]{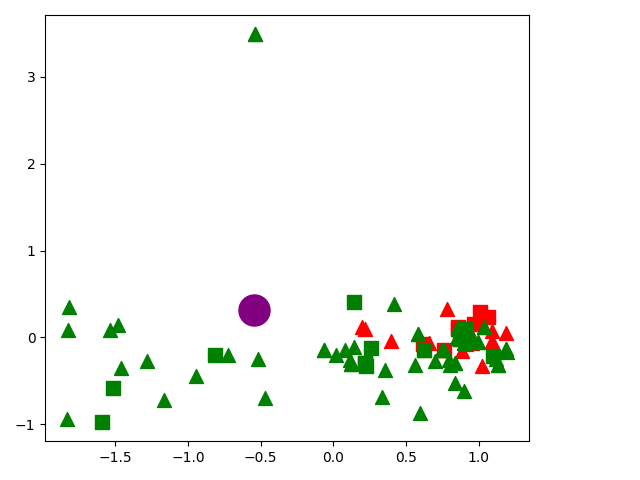}
        \end{subfigure}
         \begin{subfigure}[b]{0.138\textwidth}
         \centering
         \includegraphics[height=2.2cm, width=2.5cm]{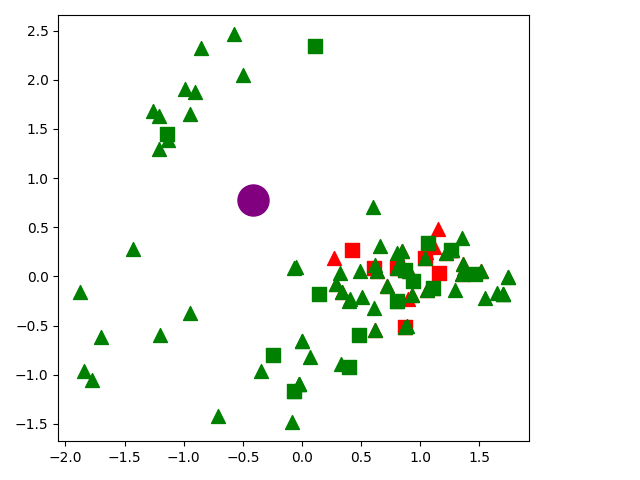}
        \end{subfigure}
         \begin{subfigure}[b]{0.138\textwidth}
         \centering
         \includegraphics[height=2.2cm, width=2.5cm]{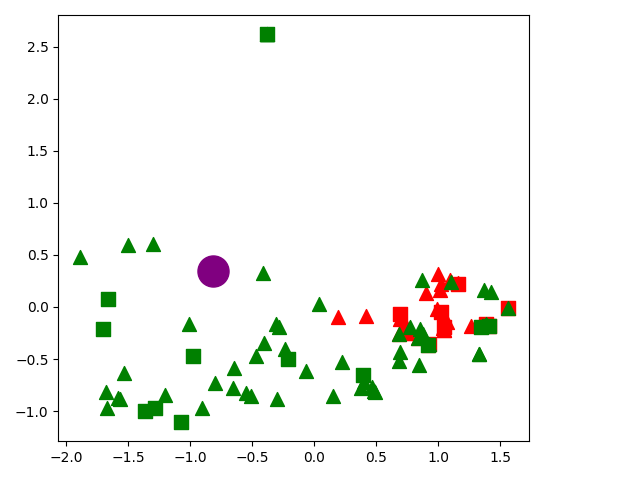}
        \end{subfigure}
         \begin{subfigure}[b]{0.138\textwidth}
         \centering
         \includegraphics[height=2.2cm, width=2.5cm]{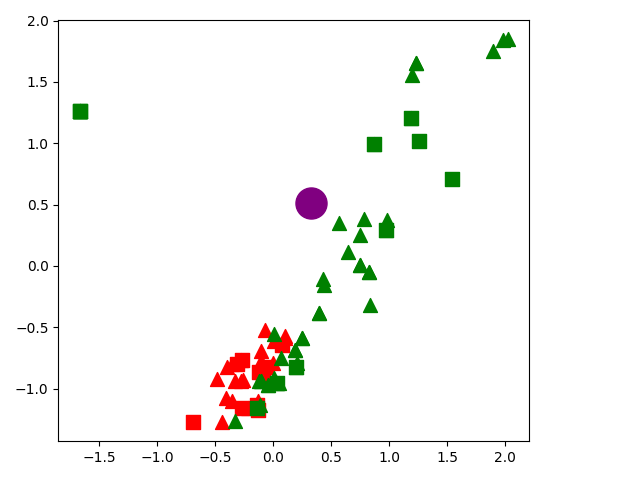}
        \end{subfigure}
         \begin{subfigure}[b]{0.138\textwidth}
         \centering
         \includegraphics[height=2.2cm, width=2.5cm]{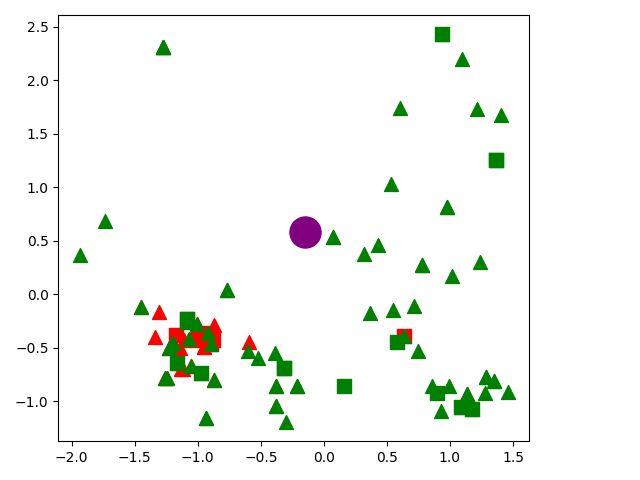}
        \end{subfigure}
 \caption{
 ICA on the bge-base-env-v1.5 embeddings for the COVID dataset, which contains $50$ input queries.
 Each figure corresponds to one of the input queries.
 The purple circles represent the queries. 
 The red squares represent the top 5 contexts ranked using cosine similarity, and the red triangles represent the corresponding hypothetical queries.
 The green squares represent the top 5 contexts ranked using our method, and the green triangles represent the corresponding hypothetical queries. 
 } 
 \label{fig:ica}
\end{figure*}

It can be observed from Fig.\ref{fig:ica} that the contexts ranked by cosine similarity tend to cluster near the input query in the embedding space. 
In contrast, the contexts ranked by HyQE and their corresponding hypothetical queries are more scattered.
This suggests that, in the embedding space, the queries are not necessarily adjacent to the contexts that provide answers to them. 
Our experimental results in Table.\ref{tab:main_result} show that the ranking produced by our HyQE has a higher NDCG@10 value than that of cosine similarity.
Therefore, both the ICA visualization and the evaluation results support our proposition that cosine similarity should be applied only when comparing queries with queries to ensure better preservation of the causal structure and to avoid spurious correlations.

\section{Additional Implementation Details} \label{sec:app_1}
In our implementation, we have used Mistral-7b-instruct-v0.2, GPT-3.5-turbo, and GPT-4o to generate hypothetical queries.

For Mistral-7b-instruct-v0.2, we use the pre-trained model. 
We set the context window size as $3900$, and the maximum number of outputs as $1024$. 
We also use an instruction prompt as shown in Fig.\ref{fig:instr} to wrap the prompt in Fig.\ref{fig:prompt}. 

\begin{figure}[!ht]
\begin{lstlisting} 
<s>[INST]\nYou are an AI assistant. Here are some rules you always follow:
- Generate human readable output, avoid creating output with gibberish text.
- Don't plainly replicate the given instruction.
- Generate only the requested output, don't include any other language before or after the requested output.
- Never say thank you, that you are happy to help, that you are an AI agent, etc. Just answer directly.
- Generate professional language typically used in business documents in North America.
- Never generate offensive or foul language.

The user prompt is as follows:\n\n\{prompt}[/INST]</s>
\end{lstlisting}

  \caption{Instruction Prompt for Mistral-7b-instruct-v0.2. `\{prompt\}' is the placeholder for the prompt shown in Fig.\ref{fig:prompt}.}
  \label{fig:instr}
\end{figure}

\begin{figure}[!ht]
 \begin{subfigure}[b]{1\textwidth}
         \centering
         \includegraphics[width=15cm]{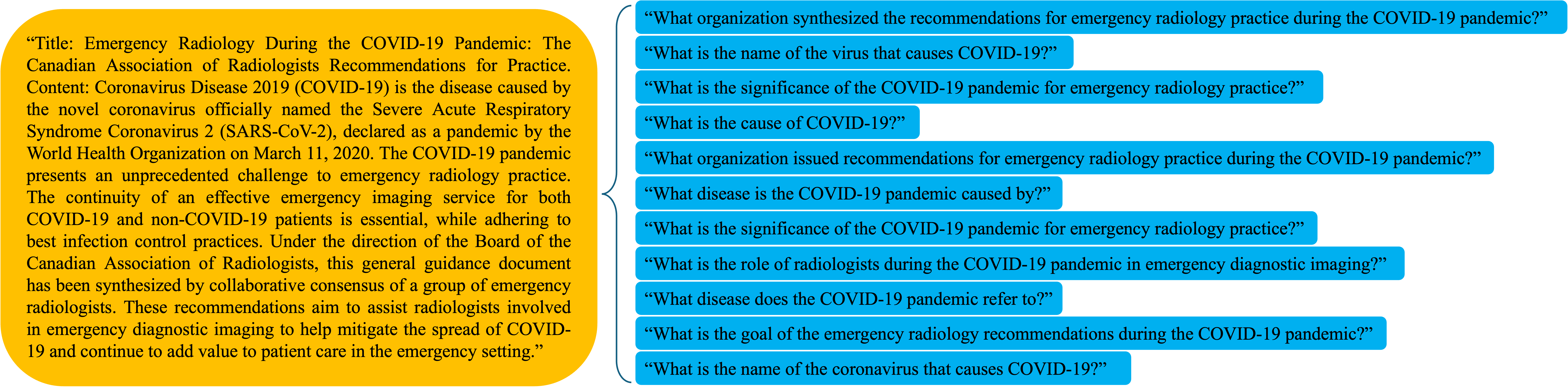}
     \end{subfigure} 
     \
     
    \begin{subfigure}[b]{1\textwidth}
         \centering
         \includegraphics[width=14cm]{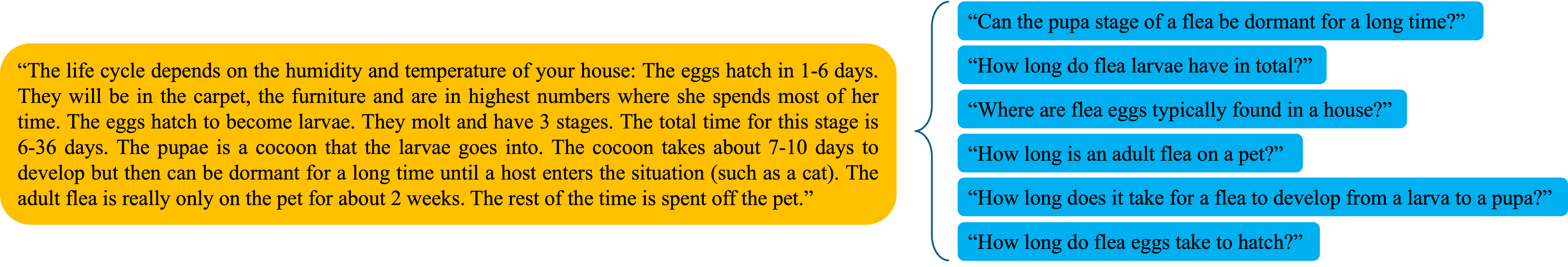}
     \end{subfigure} 
    \
    
    \begin{subfigure}[b]{1\textwidth}
         \centering
         \includegraphics[width=13cm]{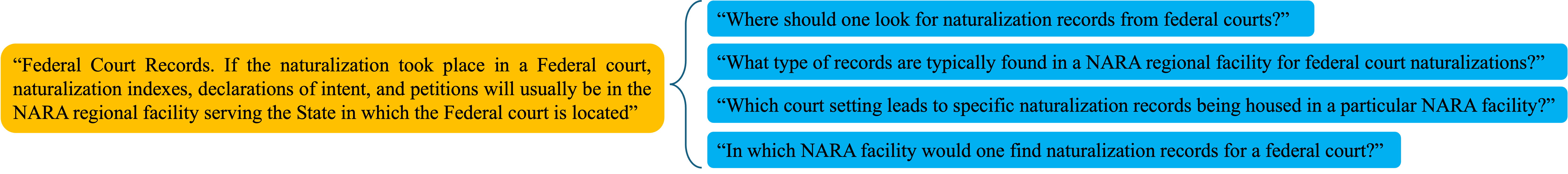}
     \end{subfigure}
\caption{
Contexts and the corresponding hypothetical queries generated by Mistral-7b-instruct-v0.2.
The contexts are in the yellow bubble.
The hypothetical queries are in the blue bubbles.
}
  \label{fig:mistral_exp}
\end{figure}

We show examples of the hypothetical queries generated by Mistral-7b-instruct-v0.2 in Fig.\ref{fig:mistral_exp}.

\begin{figure}[!ht]
 \begin{subfigure}[b]{1\textwidth}
         \centering
         \includegraphics[width=15cm]{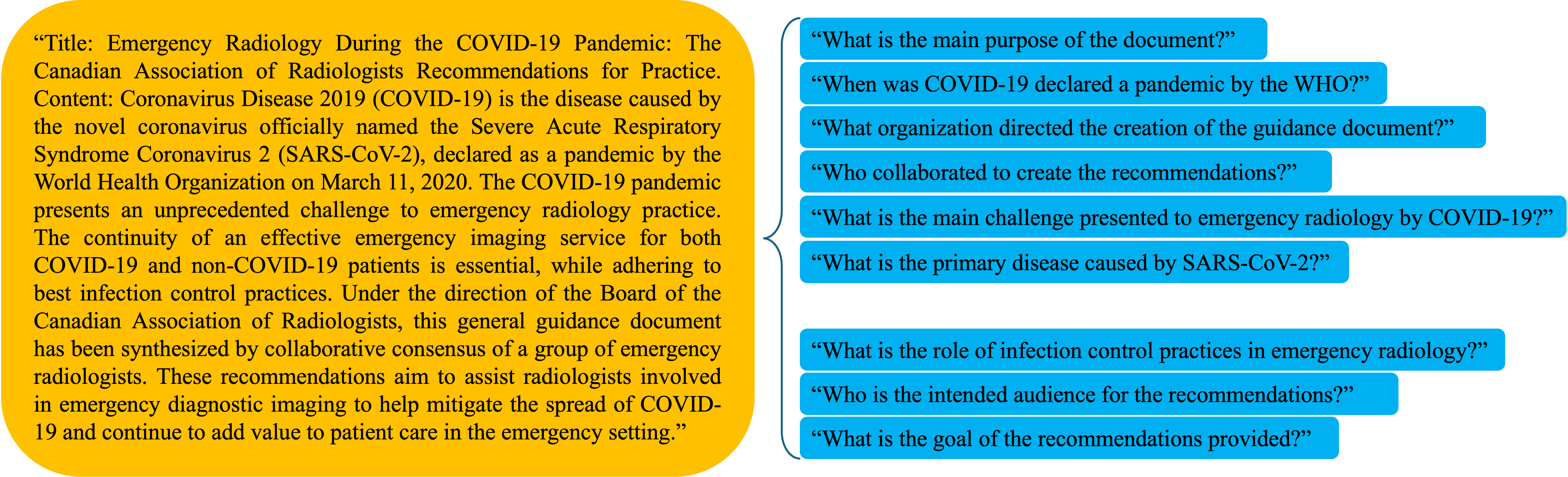}
     \end{subfigure} 
     \
     
    \begin{subfigure}[b]{1\textwidth}
         \centering
         \includegraphics[width=14cm]{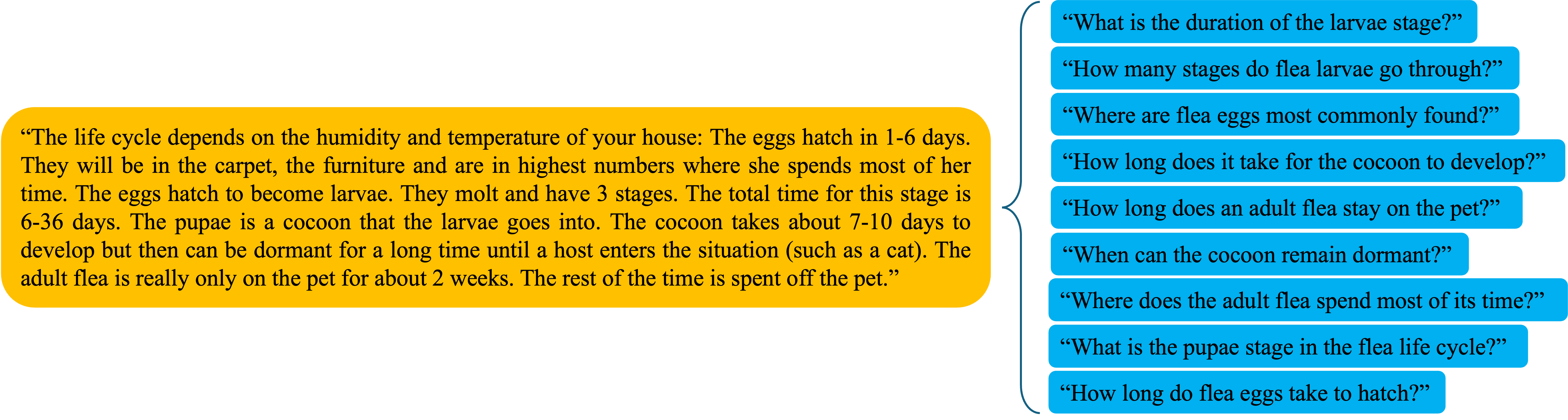}
     \end{subfigure} 
    \
    
    \begin{subfigure}[b]{1\textwidth}
         \centering
         \includegraphics[width=13cm]{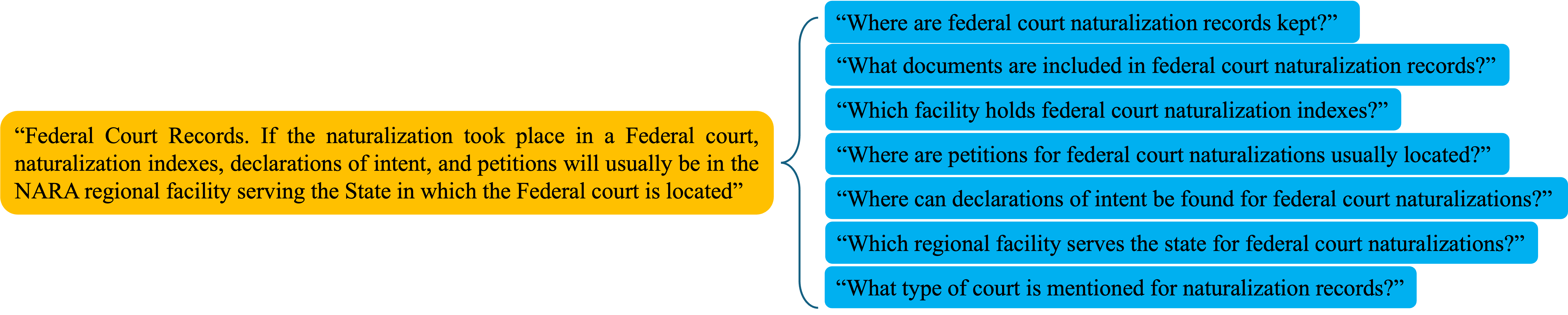}
     \end{subfigure}
\caption{
Contexts and the corresponding hypothetical queries generated by GPT-4o.
The contexts are in the yellow bubble.
The hypothetical queries are in the blue bubbles.
}
  \label{fig:gpt_exp}
\end{figure}

For GPT-3.5-turbo and GPT-4o, we send the following message to OpenAI API with the parameters $\mathtt{temperature=0.1, top\_k=1}$ and $\mathtt{n=1}$ in the request.
For the same contexts in Fig.\ref{fig:mistral_exp}, GPT-4o generates the queries as shown in Fig.\ref{fig:gpt_exp}.

\begin{figure}[!ht]
\begin{lstlisting} 
{
    "role": "system",
    "content": "
        You are an AI assistant. Here are some rules you always follow:
        - Generate human readable output, avoid creating output with gibberish text.
        - Don't plainly replicate the given instruction.
        - Generate only the requested output, don't include any other language before or after the requested output.
        - Never say thank you, that you are happy to help, that you are an AI agent, etc. Just answer directly.
        - Generate professional language typically used in business documents in North America.
        - Never generate offensive or foul language,
        "
},
{
    "role": "user",
    "content": {prompt},
}
\end{lstlisting}

  \caption{Messages sent to OpenAI API. `\{prompt\}' is the placeholder for the prompt shown in Fig.\ref{fig:prompt}.}
  \label{fig:openai}
\end{figure}

We mentioned in Section \ref{sec:exp} that we use a different prompt from that in Fig.\ref{fig:prompt} for the Touche dataset. 
The prompt is shown in Fig.\ref{fig:prompt_touche}.
We designed this prompt because each query in this dataset is about the topic of an argument, and the contexts record the dialogues in the argument, which may deviate from the topic.
An example is provided in Fig.\ref{fig:diagram}.
\begin{figure}[!ht]
\begin{lstlisting} 
Which topics could the 'Content' section of the following passage be arguing about.
If the 'Content' section provides no meaningful argument, respond with a single 'No content'.

```<passage>
{context}
</passage>``` 

Topics are questions.
Each question must be very short, different, and be written on separate lines.
Do not mention the passage itself or the author of the passage...
\end{lstlisting}
\caption{Prompt designed for the Touche2020 dataset. `\{context\}' is the placeholder for the context.}
\label{fig:prompt_touche}
\end{figure}

In Table \ref{tab:lambda} we show the hyperparameter $\lambda$ we set for each embedding model to obtain the results in Table \ref{tab:main_result}.
Note that for bge-base-env-v1.5, we use a much smaller $\lambda$ than other models because we do not normalize the product between the embeddings of the input queries and hypothetical queries but normalize the product between the embeddings of the queries and contexts. 
In this way, we obtain better and more stable results than those when we normalize all the inner products.
\begin{table*}[!ht]
\centering
\resizebox{0.98\columnwidth}{!}{
\begin{tabular}{ c | c | c | c | c | c } 
\toprule
    & contriever & bge-base-en-v1.5 & E5-large-v2 & nomic-embed-text-v1.5 & text-embedding-3-large \\
 \hline
$\lambda$  & $2.0$ & $0.03$ & $0.5$ & $0.5$ & $0.3$ \\
\bottomrule
\end{tabular}
}
\caption{Hyperparameter $\lambda$ used for each embedding model to produce results in Table \ref{tab:main_result}.}
\label{tab:lambda}
\end{table*}

Next, we show the derivation of ELBO in Eq.\ref{eq:elbo}.
\begin{eqnarray}
&&D_{KL}(p_q(c)|| p(c|q))\nonumber\\
&=&\mathbb{E}_{c\sim p_q(c)}[\log p_q(c) - \log p(c|q)]\nonumber\\
&=&\mathbb{E}_{c\sim p_q(c)}[\log p_q(c) - \log \frac{p(q|c)p(c)}{p(q)}]\nonumber\\
&=& \mathbb{E}_{c\sim p_q(c)}[\log p_q(c) - \log p(c)] - \mathbb{E}_{c\sim p_q(c)}[\log {p(q|c)}] + \mathbb{E}_{c\sim p_q(c)}[\log p(q)]\nonumber\\
&=& D_{KL}(p_q(c)||p(c)) - \mathbb{E}_{c\sim p_q(c)}[\log p(q|c)] + \log p(q)\nonumber\\
&\leq &ELBO\nonumber
\end{eqnarray}

\section{Additional Experimental Results}
\label{sec:app_2}

\begin{figure*}[!tp]
     \centering
     \begin{subfigure}[b]{0.32\textwidth}
         \centering
         \includegraphics[width=\textwidth]{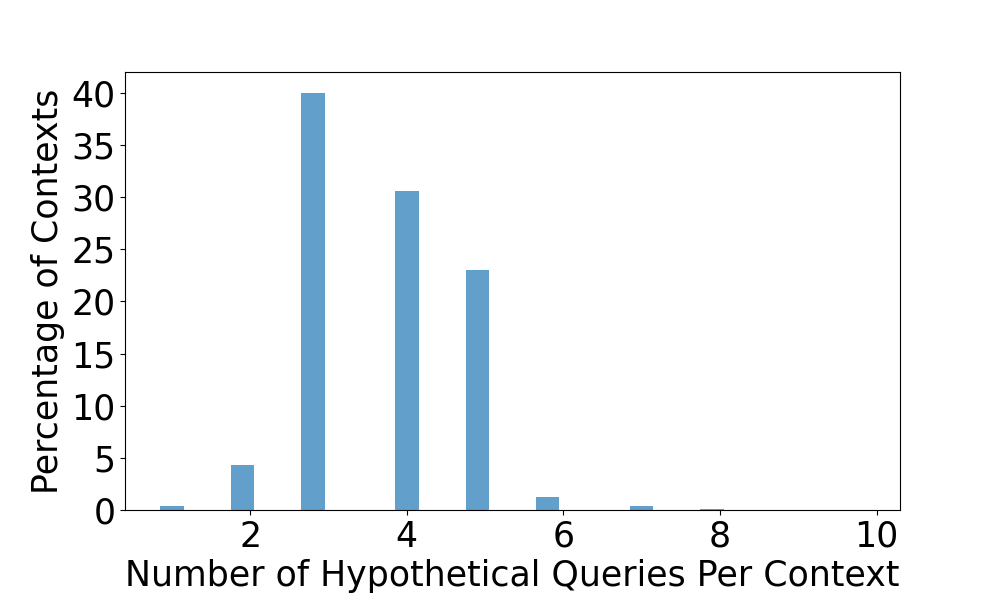}
         \caption{Gpt-3.5-Turbo on DL19}
     \end{subfigure}
     \hfill
     \begin{subfigure}[b]{0.32\textwidth}
         \centering
         \includegraphics[width=\textwidth]{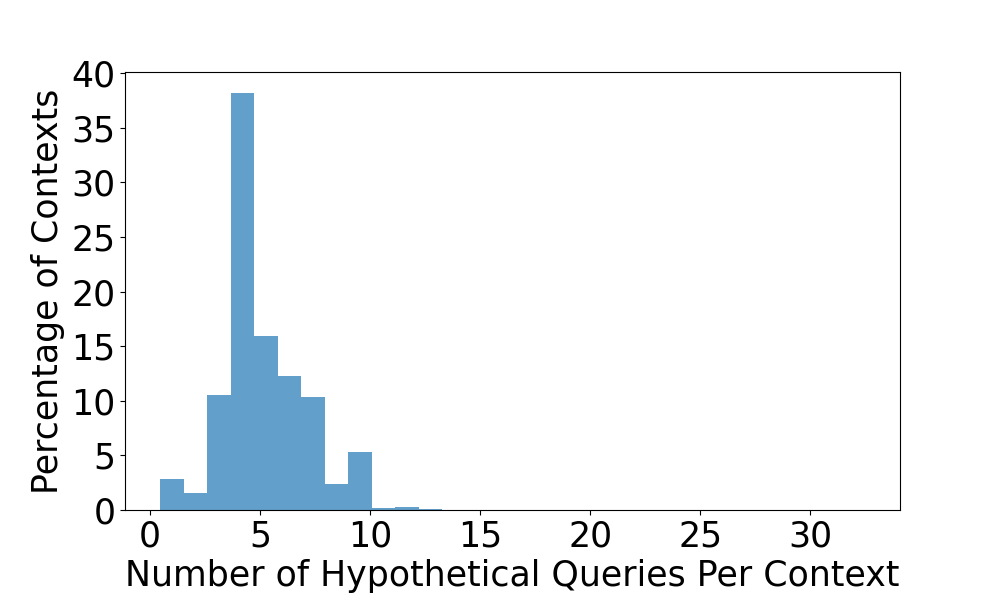}
         \caption{Gpt-4o on DL19}
     \end{subfigure}
     \hfill
     \begin{subfigure}[b]{0.32\textwidth}
         \centering
         \includegraphics[width=\textwidth]{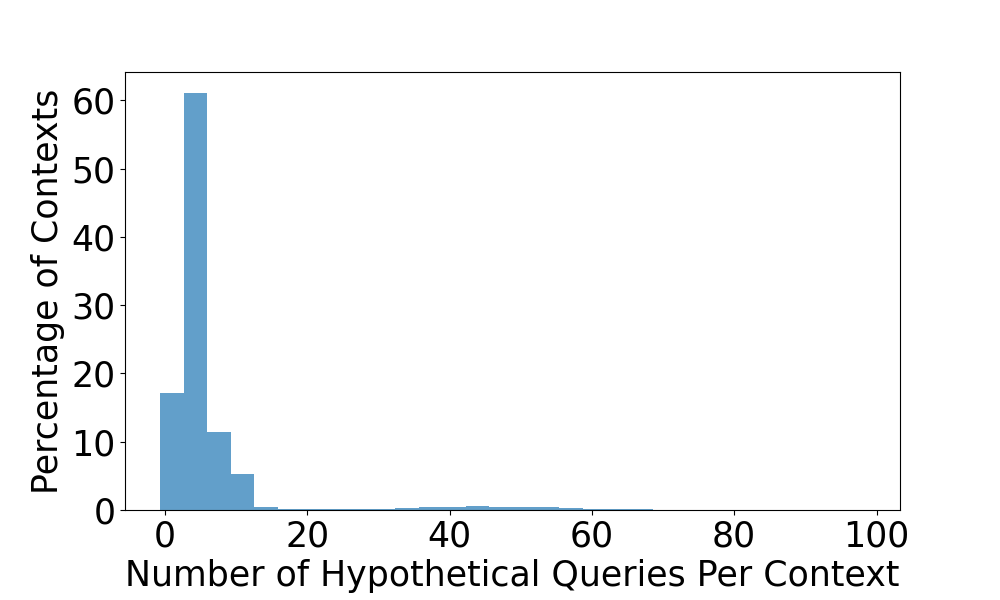}
         \caption{Mistral-7b-instruct on DL19}
     \end{subfigure} 
    \begin{subfigure}[b]{0.32\textwidth}
         \centering
         \includegraphics[width=\textwidth]{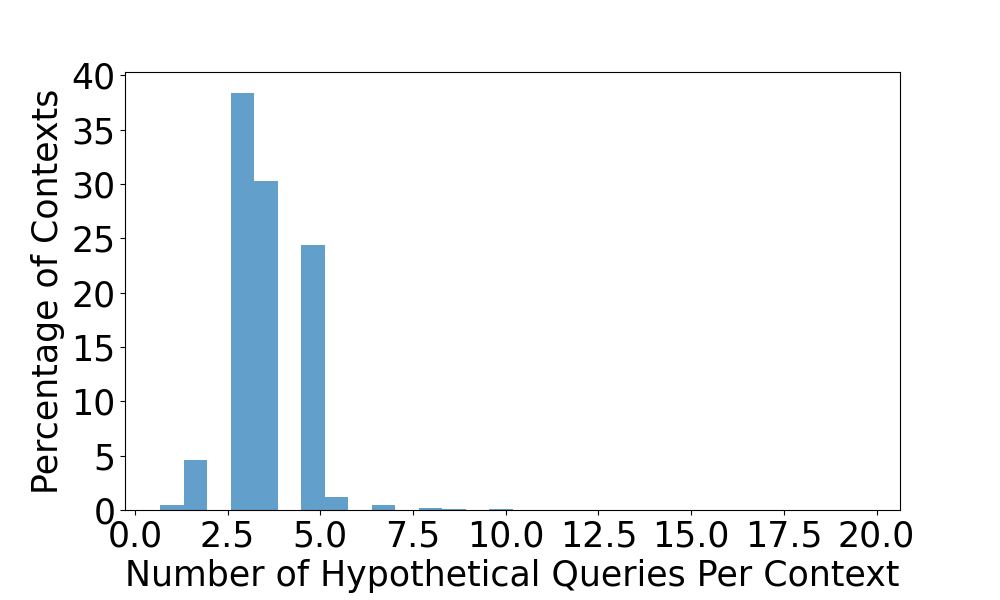}
         \caption{Gpt-3.5-Turbo on DL20}
     \end{subfigure} 
    \begin{subfigure}[b]{0.32\textwidth}
         \centering
         \includegraphics[width=\textwidth]{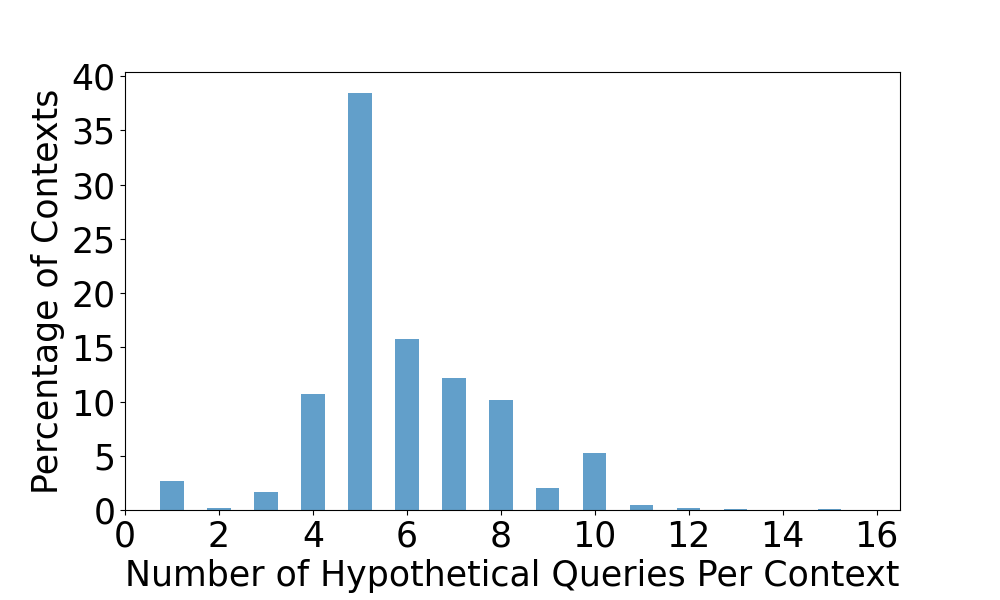}
         \caption{Gpt-4o on DL20}
     \end{subfigure} 
     \begin{subfigure}[b]{0.32\textwidth}
         \centering
         \includegraphics[width=\textwidth]{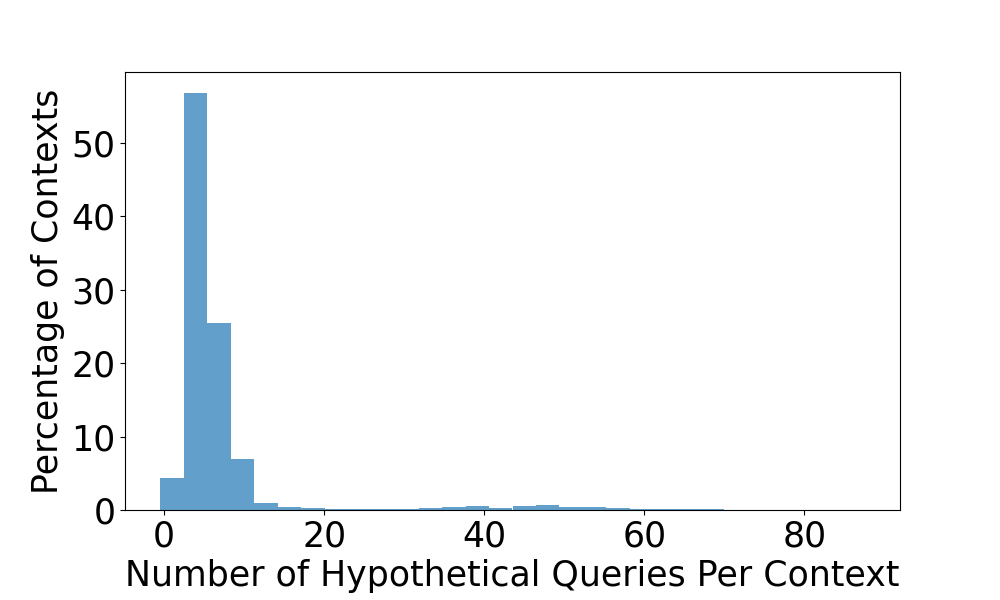}
         \caption{Mistral-7b-instruct on DL20}
     \end{subfigure} 
    \begin{subfigure}[b]{0.32\textwidth}
         \centering
         \includegraphics[width=\textwidth]{assets/plots/num_queries/gpt-3.5-turbo_covid_num_queries.png}
         \caption{Gpt-3.5-Turbo on COVID}
     \end{subfigure} 
    \begin{subfigure}[b]{0.32\textwidth}
         \centering
         \includegraphics[width=\textwidth]{assets/plots/num_queries/gpt-4o_covid_num_queries.png}
         \caption{Gpt-4o on COVID}
     \end{subfigure} 
     \begin{subfigure}[b]{0.32\textwidth}
         \centering
         \includegraphics[width=\textwidth]{assets/plots/num_queries/Mistral_covid_num_queries.png}
         \caption{Mistral-7b-instruct on COVID}
     \end{subfigure} 
    \begin{subfigure}[b]{0.32\textwidth}
         \centering
         \includegraphics[width=\textwidth]{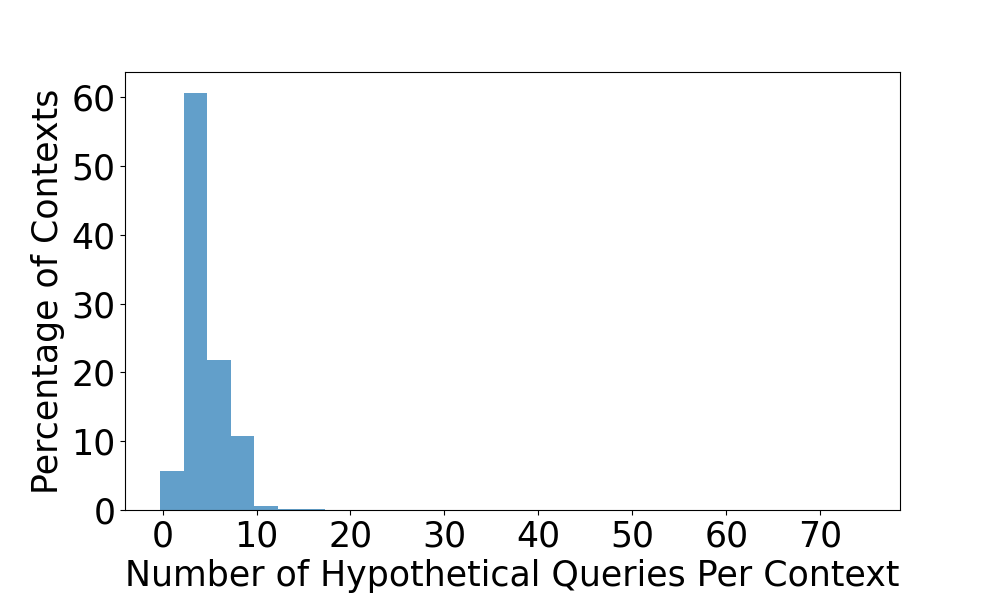}
         \caption{Gpt-3.5-Turbo on NEWS}
     \end{subfigure} 
    \begin{subfigure}[b]{0.32\textwidth}
         \centering
         \includegraphics[width=\textwidth]{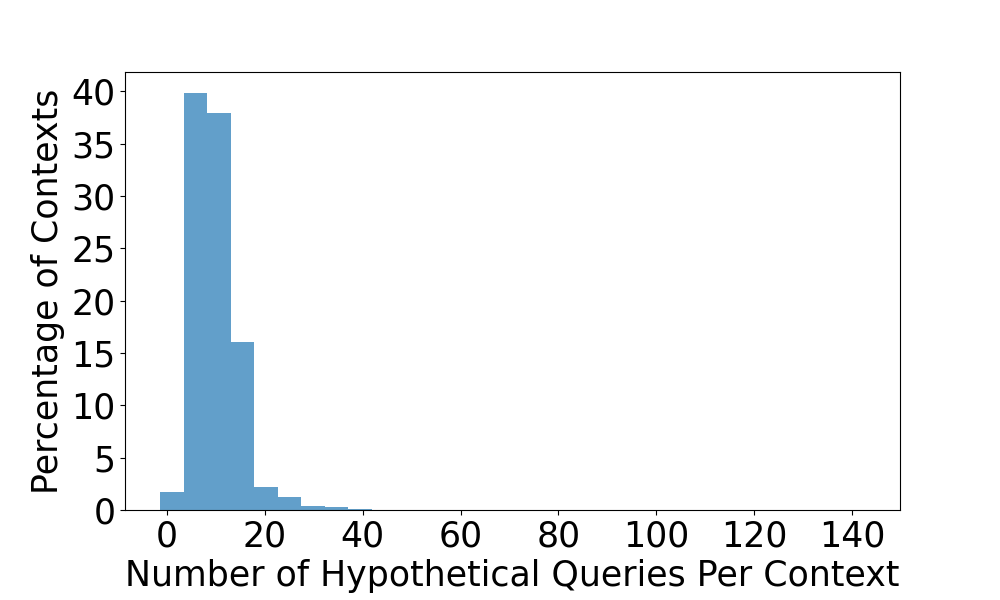}
         \caption{Gpt-4o on NEWS}
     \end{subfigure} 
     \begin{subfigure}[b]{0.32\textwidth}
         \centering
         \includegraphics[width=\textwidth]{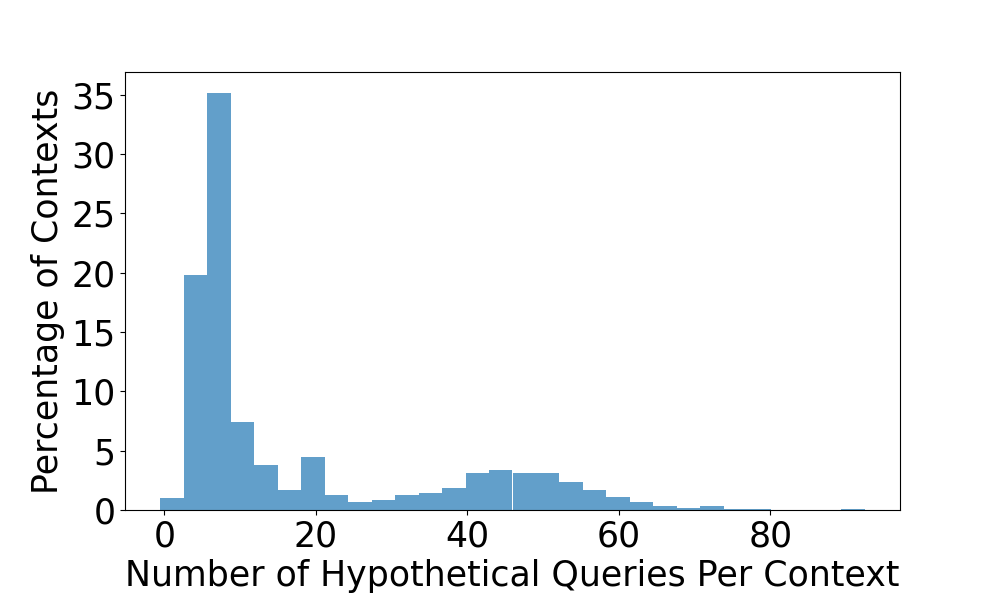}
         \caption{Mistral-7b-instruct on NEWS}
     \end{subfigure} 
    \begin{subfigure}[b]{0.32\textwidth}
         \centering
         \includegraphics[width=\textwidth]{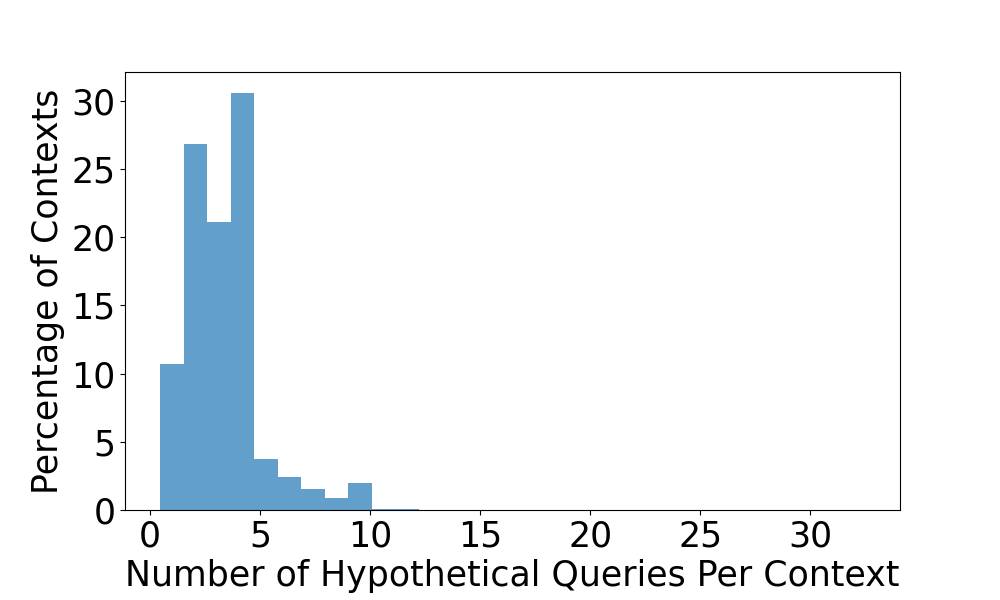}
         \caption{Gpt-3.5-Turbo on TOUCHE}
     \end{subfigure} 
    \begin{subfigure}[b]{0.32\textwidth}
         \centering
         \includegraphics[width=\textwidth]{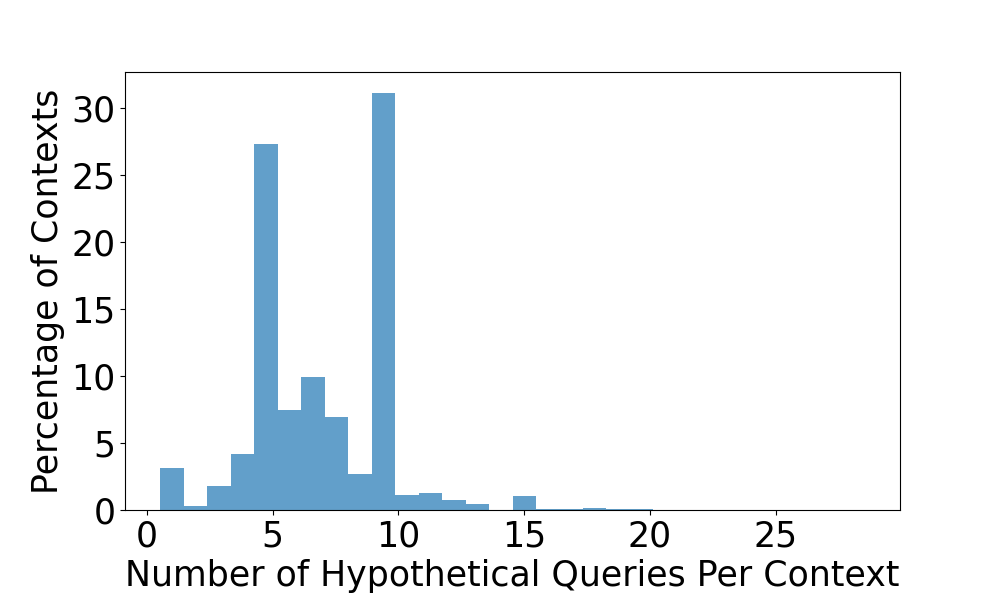}
         \caption{Gpt-4o on TOUCHE}
     \end{subfigure} 
     \begin{subfigure}[b]{0.32\textwidth}
         \centering
         \includegraphics[width=\textwidth]{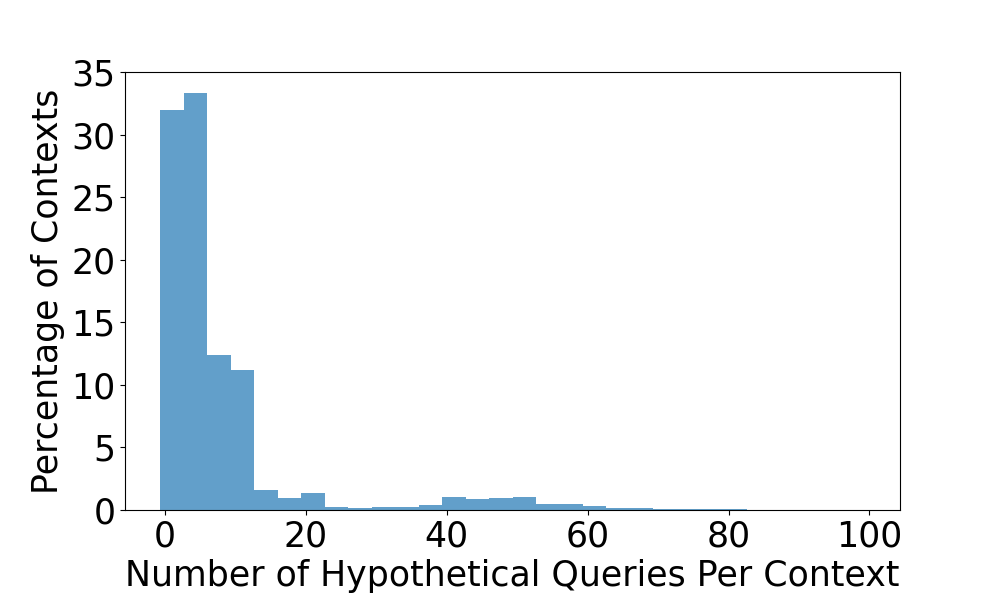}
        \caption{Mistral-7b-instruct on TOUCHE}
     \end{subfigure}
    \caption{The x-axis indicates the number of hypothetical queries generated for a context. The y-axis indicates the percentage of contexts in the dataset.
    }
    \label{fig:num_queries}
\end{figure*}

In Fig.\ref{fig:num_queries} we show the statistics of the hypothetical queries generated for each context in the benchmark datasets. 
For most of the contexts across the datasets, the LLM only generates less then $10$ hypothetical queries.
Furthermore, Table \ref{tab:avg_token_len} shows that the average token lengths of the hypothetical queries generated are around $10$ for most benchmarks. 
This attributes to our designed prompts requiring the LLMs to generate atomic queries for the contexts.

\begin{table}
\centering
\resizebox{0.5\columnwidth}{!}{
\begin{tabular}{m{7.7em} | m{2.3em}  m{2.3em} m{2.3em} m{2.3em} m{2.3em}} 
\toprule
HyQE Model  & DL19 & DL20 & COVID & NEWS & Touche \\
\hline
{Gpt-3.5-Turbo} & 11.30	&  11.41 & 	13.75 &  16.87 &  10.35 \\
\hline
Gpt-4o &	8.80	& 8.90	& 11.45	 & 9.17	 & 7.14 \\
\hline
Mistral-7b-Instruct &	12.83 &	13.24 &	16.43 &	14.80 &	22.48 \\
 \hline
\bottomrule
\end{tabular}
}
\caption{ 
Average number of tokens for each hypothetical query generated in each dataset.
}
\label{tab:avg_token_len}
\end{table}

\begin{figure}[htp!]
    \begin{subfigure}[t]{0.96\textwidth}
         \includegraphics[height=0.64cm, width=15cm]{assets/legend.png}
    \end{subfigure}  
    %
    \\
    \begin{subfigure}[b]{0.32\textwidth}
         \centering
         \includegraphics[height=3.8cm, width=5.4cm]{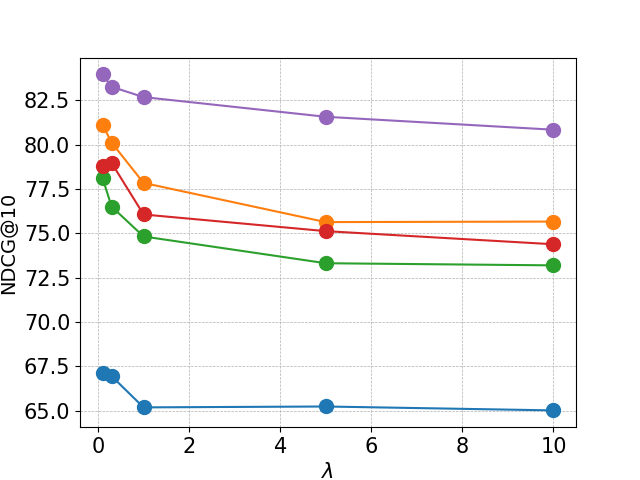}
         \caption{COVID}
     \end{subfigure} 
    \begin{subfigure}[b]{0.32\textwidth}
         \centering
         \includegraphics[height=3.8cm, width=5.4cm]{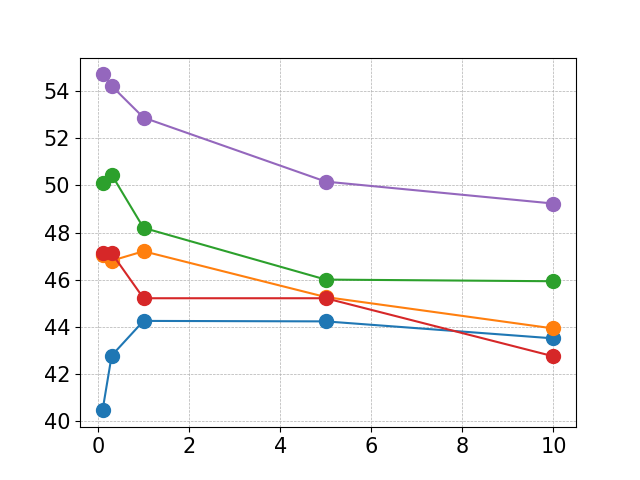}
         \caption{NEWS}
     \end{subfigure} 
    \begin{subfigure}[b]{0.32\textwidth}
         \centering
         \includegraphics[height=3.8cm, width=5.4cm]{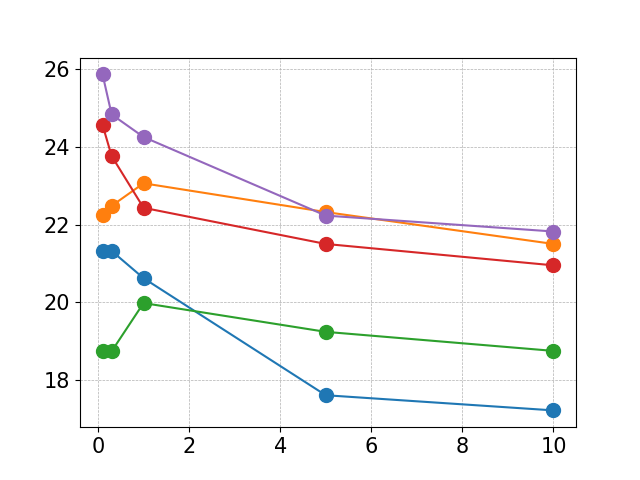}
         \caption{Touche}
     \end{subfigure}
        \caption{NDCG@10 changes with $\lambda$.
        }
    \label{fig:par_full}
\end{figure}

In Section \ref{sec:exp}, we have shown how changing the hyperparameter $\lambda$ affects HyQE on the DL19 and DL20 datasets. 
We now show the results on $3$ other datasets in Fig.\ref{fig:par_full}.
Most of the results align with those in the main text, suggesting choosing a small $\lambda$ for all models except for contriever.

We have also tried to use different embedding models for retrieval and ranking. 
As shown in Table \ref{tab:other_result}, the results align with those reported in the main text, indicating that HyQE can enhance the ranking quality.

\begin{table*}[!ht]
\centering
\resizebox{0.9\columnwidth}{!}{%
\begin{tabular}{m{6.5em} | m{9.2em} | m{6em} | m{2.3em}  m{2.3em} m{2.3em} m{2.3em}}
\toprule
Retrieval Model & Embedding Model & HyQE Model  & DL19 & DL20 & COVID & NEWS \\
\hline
\multirow{6}*{contriever}               & \multirow{2}*{bge-base-en-v1.5}       &  -     & 65.52        & 62.29 & 51.60       & 42.59     \\  
                                                                                \cline{3-7}
                                        &                                       &  GPT-3.5-turbo & {\color{blue}66.16}  & 62.15 & {\color{blue}53.80} & {\color{blue}42.69} \\  
                                        \cline{2-7}
                                        & \multirow{2}*{E5-large-v2}            &  -     & 66.24 & 65.20 & 47.08 & 46.72\\  
                                                                                \cline{3-7}
                                        &                                       &  GPT-3.5-turbo & {\color{blue}66.44} & 64.94 & {\color{blue}51.51} & {\color{blue}47.17}  \\  
                                        \cline{2-7}
                                        & \multirow{2}*{nomic-embed-text-v1.5}  &  -     & 63.27 & 60.07 & 54.07 & 43.34 \\ 
                                                                                \cline{3-7}
                                        &                                       &  GPT-3.5-turbo & {\color{blue}64.52} & {\color{blue}62.09} & 53.39 & {\color{blue}44.20} \\ 
\hline
\multirow{6}*{bge-base-en-v1.5}         & \multirow{2}*{contriever}             &  -     & 52.78 & 51.10 & 63.57 & 40.17 \\   
                                                                                \cline{3-7}
                                        &                                       &  GPT-3.5-turbo & {\color{blue}59.56} & {\color{blue}56.73} & {\color{blue}73.62} & {\color{blue}45.47}  \\  
                                        \cline{2-7}
                                        & \multirow{2}*{E5-large-v2}            &  -     & 69.48 & 71.01 & 66.19 & 48.10 \\  
                                                                                \cline{3-7}
                                        &                                       &  GPT-3.5-turbo & {\color{blue}71.92} & {\color{blue}71.36} & {\color{blue}77.62} & {\color{blue}48.41} \\  
                                        \cline{2-7}
                                        & \multirow{2}*{nomic-embed-text-v1.5}  &  -     & 68.20 & 65.61 & 77.25 & 43.60 \\
                                                                                \cline{3-7}
                                        &                                       &  GPT-3.5-turbo & {\color{blue}71.28} & {\color{blue}67.20} & {\color{blue}77.69} & {\color{blue}44.30} \\ 
\bottomrule
\end{tabular}
}
\caption{
NDCG@10 results produced by different combinations of embedding models across various datasets. The `$-$’ sign indicates that the results in the associated row are generated without HyQE. The blue color highlights that using HyQE for ranking results in a higher NDCG@10 value compared to not using HyQE for the combination of embedding models and dataset. 
}
\label{tab:other_result}
\end{table*}

In addition to using different LLMs to generate the hypothetical queries, we also vary the temperatures of the LLM during the generation. 
Higher temperatures can make the LLM's response more random. 
In Table \ref{tab:temp}, we utilized contriever for retrieval and embedding and used GPT-3.5-turbo for query generation under different temperatures. 
The results show that the ranking performances of our approach is not significantly impacted by the temperature.

\begin{table}
\centering
\resizebox{0.5\columnwidth}{!}
{
\begin{tabular}{m{5em} | m{2.3em}  m{2.3em}   m{2.3em}  m{2.3em}  m{2.3em}} 
\toprule
Temperature &  DL19 & DL20 & COVID & NEWS & Touche \\
\hline
0.1	    & {53.19}   & {50.04}  & {35.06}  & {42.33} & {21.02}\\ 
                                          
\hline
0.5     & {53.05}   & {49.50}  & {35.70}  & {40.96} & {20.43}\\                                           
\hline
1.0     & {52.56}   & {49.74}  & {36.10}  & {41.98} & {21.91}\\ 
\bottomrule
\end{tabular}
 }
\caption{  
NDCG@10 results produced by using contriever for retrieval and embedding, and using GPT-3.5-turbo for query generation under different temperatures. 
}
\label{tab:temp}
\end{table}

In Algorithm \ref{algo:hyqe}, the parameter $K$ in the candidate context set $C_{q, K}$ functions can also be considered as a hyperparameter.
Setting a small value for $K$ limits the range of contexts to be ranked, resulting in fewer calls to the LLM. 
Conversely, a large value of $K$ allows for low-rank but potentially highly relevant contexts to be re-ranked. 
However, this increases the number of calls to the LLM and the risk of erroneously assigning a high rank to a low-relevant context. 
In Section \ref{sec:exp}, the results are obtained with $K$ set to 30. 
In Table \ref{tab:k}, we show how the performance of HyQE changes with the value of $K$.
Compared with Table \ref{tab:main_result}, the results for $K=20$ and $K=30$ are close to each other.
\begin{table*}[!ht]
\centering
\resizebox{1\columnwidth}{!}{%
\begin{tabular}{m{6.5em} | m{9.6em} | m{7.7em}  | m{2.3em} | m{2.3em}  m{2.3em} m{2.3em} m{2.3em}} 
\toprule
Retrieval Model & Embedding Model & HyQE Model & K  & DL19 & DL20 & COVID & NEWS \\
\hline
\multirow{16}*{contriever}              & \multirow{4}*{contriever}               & \multirow{2}*{GPT-3.5-turbo}        &  10  & 46.35         & 43.56       & 28.84        &  36.74    \\  
                                                                                                                        \cline{4-8}
                                        &                                         &                                     &  20 & 51.38         & 48.59        & 32.82        &  41.33    \\ 
                                                                                  \cline{3-8}
                                        &                                         & \multirow{2}*{Mistral-7b-instruct}  &  10 & 46.14         & 42.94        & 28.63        &  36.24   \\  
                                                                                                                        \cline{4-8}
                                        &                                         &                                     &  20 & 50.85         & 48.16        & 32.90        &  40.18  \\ 
                                        \cline{2-8}                             
                                        & \multirow{4}*{bge-base-en-v1.5}         & \multirow{2}*{GPT-3.5-turbo}        &  10 & 66.55         & 61.84        & 52.66        &  42.09     \\  
                                                                                                                        \cline{4-8}
                                        &                                         &                                     &  20 & 66.16         & 62.15        & 53.80        &  42.69   \\ 
                                        \cline{3-8}
                                        &                                         & \multirow{2}*{Mistral-7b-instruct}  &  10 & 66.58         & 61.94        & 52.62        &  42.31   \\  
                                                                                                                        \cline{4-8}
                                        &                                         &                                     &  20 & 65.89         & 62.40        & 53.93        &  42.99  \\ 
                                        \cline{2-8}    
                                        
                                        & \multirow{4}*{E5-large-v2}              & \multirow{2}*{GPT-3.5-turbo}        &  10 & 66.55         & 64.85        & 27.32        &  34.84   \\  
                                                                                                                        \cline{4-8}
                                            &                                         &                                 &  20 & 66.48         & 64.98        & 27.32        &  34.84  \\
                                                                                  \cline{3-8}
                                        &                                         & \multirow{2}*{Mistral-7b-instruct}  &  10 & 66.41         & 64.84        & 48.38        &  46.08   \\  
                                                                                                                        \cline{4-8}
                                        &                                         &                                     &  20 & 65.67         & 65.09        & 52.78        &  46.28  \\ 
                                        \cline{2-8}
                                        & \multirow{4}*{nomic-embed-text-v1.5}    & \multirow{2}*{GPT-3.5-turbo}        &  10 & 62.12         & 61.77        & 53.92        &  44.83   \\  
                                                                                                                        \cline{4-8}
                                        &                                         &                                     &  20 & 64.11         & 62.08        & 53.26        &  43.86  \\
                                                                                  \cline{3-8}   
                                        &                                         & \multirow{2}*{Mistral-7b-instruct}  &  10 & 64.48         & 61.54        & 55.46        &  43.09   \\  
                                                                                                                        \cline{4-8}
                                        &                                         &                                     &  20 & 63.47         & 63.69        & 54.71        &  44.22  \\ 
\bottomrule
\end{tabular}
}
\caption{
NDCG@10 results produced by embedding models and hypothetical query generators (LLMs) across various datasets. The values in the $K$ column indicates HyQE is used to re-rank the top-$K$ contexts ordered by the embedding model.
}
\label{tab:k}
\end{table*}

\end{document}